 \documentclass[10pt,preprint]{aastex}  %compact preprint style (MJA)
 \usepackage{natbib}
\def\ang{\AA}

\def\gapprox{\lower.4ex\hbox{$\;\buildrel >\over{\scriptstyle\sim}\;$}}
\def\lapprox{\lower.4ex\hbox{$\;\buildrel <\over{\scriptstyle\sim}\;$}}

\shortauthors{ASCHWANDEN 2013}
\shorttitle{Macroscopic SOC Description}

\begin{document}
%{\sl  Manuscript, accepted ... }

\title{         A Macroscopic Description of a Generalized Self-Organized 
		Criticality System: Astrophysical Applications}

\author{        Markus J. Aschwanden}

\affil{		Lockheed Martin Solar and Astrophysics Laboratory, 
                A021S, Bldg.252, 3251 Hanover St.,
                Palo Alto, CA 94304, USA;
                e-mail: aschwanden@lmsal.com}

\begin{abstract}
We suggest a generalized definition of self-organized criticality (SOC) 
systems: SOC is a critical state of a nonlinear energy dissipation 
system that is slowly and continuously driven towards a critical value 
of a system-wide instability threshold, producing scale-free, 
fractal-diffusive, and intermittent avalanches with powerlaw-like size 
distributions. We develop here a macroscopic description of SOC systems
that provides an equivalent description of the complex microscopic fine
structure, in terms of fractal-diffusive transport (FD-SOC). Quantitative
values for the size distributions of SOC parameters (length scales $L$,
time scales $T$, waiting times $\Delta t$, fluxes $F$, and fluences
or energies $E$) are derived from first 
principles, using the scale-free probability conjecture, $N(L) dL \propto 
L^{-d}$, for Euclidean space dimension $d$. We apply this model to
astrophysical SOC systems, such as lunar craters, the asteroid belt,
Saturn ring particles, magnetospheric substorms, radiation belt electrons,
solar flares, stellar flares, pulsar glitches, soft gamma-ray repeaters, 
black-hole objects, blazars, and cosmic rays. The FD-SOC model predicts 
correctly the size distributions of 8 out of these 12 astrophysical 
phenomena, and indicates non-standard scaling laws and measurement biases 
for the others.
\end{abstract}


\section{INTRODUCTION}

Although the paradigms of {\sl self-organized criticality (SOC)} systems
appear to be very intuitive and self-explaining, such as the 
self-adjusting angle of repose in Per Bak's sandpile (Bak et al.~1987), 
or the stick-slip motion of earthquakes (Gutenberg and Richer 1949),
theoreticians find it hard to establish a rigorous general definition 
of SOC systems. Part of the problem are the subtle differences between
``criticality'' in fine-tuned systems that undergo percolation or
phase transitions, such as the Ising model (Ising 1925), versus 
``self-organized criticality'' systems, which do not need any fine-tuning 
(e.g., Christensen and Moloney 2005).
A solid definition of SOC systems 
should (i) be able to make quantitative predictions that are testable
by observations, and (ii) provide discrimination criteria between
SOC and alternative transport processes occurring in complex systems
(such as random walk, branching theory, network theory, percolation, 
aggregation, or turbulence). A mathematical definition of SOC includes
{\sl``non-trivial scale invariance (with spatio-temporal correlations)
in avalanching (intermittent) systems as known from ordinary critical
phenomena, but with internal, self-organized rather than external
tuning of a control parameter (to a non-trivial value)''} (Pruessner
2012). Alternatively, we may define SOC from a more physical point of view:
{\sl SOC is a critical state of a nonlinear energy dissipation system that 
is slowly and continuously driven towards a critical value of a system-wide
instability threshold, producing scale-free, fractal-diffusive, and 
intermittent avalanches with powerlaw-like size distributions.} 
This definition applies to SOC phenomena as diverse as sandpiles, 
earthquakes, solar flares, or stockmarket fluctuations. 

The major problem is that SOC is a microscopic process in complex
systems, which cannot easily be described by macroscopic equations,
unlike entropy-related processes in classical thermodynamics. In order
to obtain insights into SOC processes, microscopic processes in
complex systems have been simulated by iterative numerical codes,
such as cellular automaton models, where a single time step is
quantified by a mathematical redistribution rule, which operates
on a microscopic level. Such SOC models are also called 
{\sl slowly-driven interaction-dominated threshold (SDIDT)} systems,
which all share some common properties, such as a large but finite 
number of degrees of freedom, a threshold for nonlinearity, a 
re-distribution rule once the local variable exceeds the threshold, 
and a continuous but slow driver (Jensen 1998; p.126; 
Pruessner 2012; p.7). Such numerical simulations produce 
powerlaw-like probability distributions of SOC parameters, which
are generally considered as a necessary (but not satisfactory)
criterion to identify SOC.

In this study we derive a macroscopic description of SOC processes
by analytical means, which are supposed to mimic the statistics of
microscopic, spatially unresolved, next-neighbor interactions in SOC 
systems. The situation is similar to classical thermodynamics, where
macroscopic parameters such as temperature, pressure, or entropy 
describe the microscopic state (e.g., the Boltzmann distribution), 
resulting from atomic collisions and other energy dissipation processes. 
The analytical approximation of complex spatial structures is accomplished
by the concept of fractals (i.e., monofractals or multi-fractals).
Our analytical framework of SOC processes includes geometric, temporal,
physical, and observable parameters, for which physical scaling laws
exist that determine the spatio-temporal evolution and the statistical
distributions. However, the main difference to classical thermodynamics
is the nonlinear nature of complex systems, while thermodynamic systems
are governed by incoherent random noise that add up in a linear way.

\section{AN ANALYTICAL MACROSCOPIC SOC MODEL}

Our analytical description of SOC models entails four different
aspects: (1) geometric parameters and geometric scaling laws;
(2) temporal parameters and spatio-temporal evolution and transport;
(3) physical scaling laws; and (4) instrument-dependent observables.
These four domains are treated separately in the following.

\subsection{The Scale-Free Probability Conjecture}

We start with geometric parameters, such as a length scale $L$,
an Euclidean area $A$, an Euclidean volume $V$, embedded in a
Euclidean space with a dimension of $d=$1, 2, or 3. Euclidean means
space-filling here, while inhomogeneous structures are described
by a fractal dimension $D_d$, which also depends on the Euclidean
dimension $d$.

SOC phenomena (like avalanches on a sandpile) can be triggered by
the infall of a single sand grain, and thus the causal consequence
of a tiny input or disturbance can have an unpredictable magnitude of
the outcome or nonlinear response of a SOC system. Henceforth,
the geometric size of a SOC avalanche can cover a considerable range
$L_1 \le L \le L_2$ from the size $L_1$ of a single sand grain to
the finite size $L_2$ of the SOC system. If only next-neighbor
interactions are allowed in a SOC system, such as in the 
Bak-Tang-Wiesenfeld (BTW) model (Bak et al.~1987), a continuous distribution of
length scales $L$ of avalanches is expected when averaged over a
long time. Naturally, small avalanches have a higher probability
to occur than large ones, because they can happen simultaneously
at different places of a sandpile, while a large system-wide avalanche
can occur only once at a time. So, we can ask the question about the
{\sl probability distribution function (PDF)}, $N(L) dL$, of 
avalanches with size $L$ to occur in a SOC system. In order to solve 
this problem, we proceed in the same way as the PDF of random processes
is derived. 

The simplest statistical distribution is obtained from
rolling dice, by enumerating all possible outcomes. The PDF of
outcomes of rolling one dice, two dice, and three dice is shown
in Fig.~1, the classical binomial distribution that approaches 
a Gaussian normal distribution (Fig.~1) for a
large number of dice, with possible outcomes of $n \le x \le 6n$
for 6-sided dice, while the PDF $N(x) dx$ is a Gaussian function
centered at $x=n (6+1)/2$.

Going to the statistical probability distributions of avalanches 
with size $L$, we use the same method by enumerating all possible
states with size $L$ that can occur in a SOC system with finite
size $L_2$. The case with an Euclidean space dimension of $d=2$
is illustrated in Fig.~2, where we use logarithmic bins with size
$x=1, 2, 4, 8, 16$. In a system with finite size $L_2=16$, one
avalanche of this maximum size $L_2$ is possible in a given time
interval, and thus $N(x=16)=1$. For a bin with half the size,
$L=L_2/2=8$ we have four possible areas with a length scale of
$x=L_2/2$, and thus $N(x=8)=4$. Proceeding to quarter bins,
$L=L_2/4$, we have 16 possible areas with size $L=L_2/4$, and thus
$N(x=4)=16=2^4$, and so forth. Obviously, the probability distribution
scales as $N(x)dx = (L_2/L)^2$ for Euclidean dimension $d=2$.
We can easily imagine the probabilities for the other Euclidean
dimensions $d=1$, which is $N(x)dx= (L_2/L)$, and for $d=3$,
which is $N(x)dx=(L_2/L)^3$. Therefore we obtain a generalized 
probability distribution of length scales $L$ according to
\begin{equation}
	N(L) dL \propto L^{-d} \ dL \ ,
\end{equation}
which we call the {\sl scale-free probability conjecture}  
(Aschwanden 2012a), being related to {\sl packing rules} (e.g., 
sphere packing, or dense packing) in geometric aggregation problems. 
A similar approach of using geometric scaling laws was also 
pioneered for earthquakes (Main and Burton 1984).
The term {\sl scale-free} is generally used to express that no special
scale is present in a statistical distribution, unlike the first moment
or center value of a Gaussian (normal) distribution, or the e-folding
value in an exponential distribution. Our scale-free probability 
does not require that all possible avalanches in a SOC system have
to occur simultaneously, or in any particular sequential order.
They just represent the expected distribution of a statistically
representative sample, similar to the rolling of dice. For instance,
using $n=10^{26}$ dice to mimic the number of atoms per $cm^3$,
there is no way to execute all possible rolls, but we expect for any
statistically representative subset of possible outcomes a Gaussian
distribution. Similarly, we expect a length distribution $N(L)$
according to Eq.~(1) for any statistically representative subset
of avalanches occurring in a SOC system. We expect that Eq.~(1) has
universal validity in SOC systems, because it is only based on
a statistical argument of random processes on all scales,
without any other constraints given by specific physical parameters 
or the dynamic behavior of a SOC system. This scale-free probability 
conjecture (Eq.~1) may also occur in other nonlinear systems, such as in 
turbulence. We may be able to discriminate between the two systems by 
the sparseness of avalanches (in slowly-driven SOC systems) and
the space-filling of structures (in turbulent media).

\subsection{Geometric Scaling Laws}

In the following we are going to derive size distributions of
SOC avalanches by using geometric scaling laws, which is a standard
approach that has been applied in a number of previous work
(e.g., Bak et al.~1988; Robinson 1994; Munoz et al.~1999;
Biham et al.~2001).

Besides the length scale $L$, other geometric parameters are the 
Euclidean area $A$ or the Euclidean volume $V$. 
The simplest definition of an area $A$
as a function of a length scale $L$ is the square-dependence,
\begin{equation}
	A \propto L^2 \ , 
\end{equation}
which applies also to circular areas, $A \propto \pi r^2$, or
more complicated solid areas, differing only by a constant factor
for self-similar geometric shapes. A direct consequence of this
simple geometric scaling law is that the statistical probability
distribution of avalanche areas is directly coupled to the
scale-free probability distribution of length scales (Eq.~1), 
and can be computed by substitution of $L(A) \propto A^{1/2}$ (Eq.~2),
into the distribution of Eq.~(1), 
$N(L)=N(L[A])=L[A]^{-d}=(A^{1/2})^{-d}=A^{-d/2}$, 
and with the derivative $dL/dA \propto A^{-1/2}$,
\begin{equation}
	N(A) dA \propto N(L[A]) \left| {dL \over dA} \right| dA 
		\propto A^{-(1+d)/2}\ dA\ .
\end{equation}
Thus we expect an area distribution $N(A)$ depending on the dimensionality
$d=2,3$ of the SOC system, 
\begin{equation}
	N(A) dA \propto A^{-\alpha_A} dA 
	\qquad {\rm where}\ \left\{
	\begin{array}{ll}
	\alpha_A=1.5 & {\rm for}\ d=2 \\
	\alpha_A=2.0 & {\rm for}\ d=3 \\
	\end{array}
        \right.	\ ,
\end{equation}
which should also have universal validity for SOC systems. In spatially
resolved astrophysical observations, such as of the Sun or magnetosphere,
a length scale $L$ or area $A$ are the only directly measurable geometric 
parameters, while a volume $V$ is generally derived from the observed 
area of a SOC event.

Similarly to the area, we can derive the geometric scaling for volumes $V$,
which simply scales with the cubic power in 3D space,
\begin{equation}
	V \propto L^3 \ , 
\end{equation}
which represents a cube, but differs only by a constant factor for
a sphere, i.e., $V=(4\pi/3) r^3$. Consequently, we can also derive
the probability distribution $N(V) dV$ of volumes $V$ directly from
the scale-free probability conjecture (Eq.~1), where the definition
of Eq.~(5) demands $d=3$. Substituting $L \propto V^{1/3}$ into
$N(L[V]) \propto L[V]^d \propto V^{-d/3}$ and the derivative
$dL/dV=V^{-2/3}$ we obtain for $d=3$, 
\begin{equation}
	N(V) dV \propto N(L[V]) \left| {dL \over dV} \right| dV 
		\propto V^{-\alpha_V}\ dV
		\propto V^{-5/3}\ dV\ .
\end{equation}
Thus, a powerlaw slope of $\alpha_V=5/3$ is predicted in 3D
Euclidean space, which applies also to the Euclidean volume of a
time-integrated SOC avalanche in lattice simulations. However,
since avalanches have a fractal geometry, it is the time-integrated
fractal volume that is equivalent to the number of active pixels 
in a lattice simulation, rather than the Euclidean volume.

Since all the assumptions made so far are universal, such as the
scale-free probability conjecture (Eq.~1) and the geometric scaling
laws $A \propto L^2$ and $V \propto L^3$, the resulting predicted
occurrence frequency distributions of $N(A)$ (Eq.~3) and $N(V)$ 
(Eq.~6) are universal too, and powerlaw functions are predicted 
from this derivation from first principles, which is consistent 
with the property of {\sl universality} in theoretical SOC definitions.

\subsection{The Fractal Geometry}

{\sl ``Fractals in nature originate from self-organized critical 
dynamical processes''} (Bak and Chen 1989). Fractal geometries
have been pioneered in the context of self-similar structures
before the advent of SOC models (Mandelbrot 1977, 1983, 1985),
and have been applied to spatio-temporal SOC structures extensively
(e.g., Bak et al.~1987, 1988; Bak and Chen 1989; Ito and Matsuzaki 1990;
Feder and Feder 1991; Rinaldo et al.~1993; Erzan et al.~1995;
Barabasi et al.~1995). Since the fractal geometry
is a postulate of SOC processes invoked by the first pioneers of SOC,
it is appropriate to approximate spatial structures of SOC avalanches
by a fractal dimension. The simplest fractal is the Hausdorff dimension
$D_d$, which is a monofractal and depends on the Euclidean space dimension 
$d=1,2,3$. The Hausdorff dimension $D_3$ for the 3D Euclidean space ($d=3$) is
\begin{equation}
	D_3 = {\log{V_f(t)} \over \log{(L)} } \ ,
\end{equation}
and analogously for the 2D Euclidean space ($d=2$), 
\begin{equation}
	D_2 = {\log{A_f(t)} \over \log{(L)} } \ ,
\end{equation}
with $A_f(t)$ and $V_f(t)$ being the fractal area and volume of a SOC 
avalanche during an instant of time $t$. These fractal dimensions can be
determined by a box-counting method, where the area fractal $D_2$ can
readily be obtained from images from the real world, while the volume 
fractal $D_3$ is generally not available unless one obtains 3D data 
(or by numerical simulations). 
 
A good approximation for the
expected fractal dimension $D_d$ is the mean value of the smallest
possible fractal dimension $D_{d,min}\approx 1$ and the largest
possible fractal dimension $D_{d,max}=d$. The minimum possible fractal
dimension is near the value of 1 because the next-neighbour interactions
in SOC avalanches require some continuity between active nodes in a
lattice simulation of a cellular automaton, while smaller fractal
dimensions $D_d < 1$ are too sparse to allow an avalanche to propagate
via next-neighbor interactions. Thus, the mean value of a fractal
dimension is expected to be (Aschwanden 2012a),
\begin{equation}
	D_d \approx {D_{d,min} + D_{d,max} \over 2} = {(1+d) \over 2} \ .
\end{equation}
Thus, we expect fractal dimensions of $D_3\approx (1+3)/2=2.0$ for
the 3D space, and $D_2\approx (1+2)/2=1.5$ for the 2D space. 
This conjecture of the mean value of the fractal dimension $D_d$ 
has been numerically tested with cellular automaton simulations for
Eucledian dimensions $d=1,2,3$ and the following mean values were
found: $D_1=1.00\pm0.00$ (Aschwanden 2012a); then $D_2=1.58\pm0.02$
(Charbonneau et al.~2001), $D_2=1.58\pm0.03$ (McIntosh et al.~2002),
$D_2=1.60\pm0.17, 1.62\pm0.18$ (Aschwanden 2012a) for the 2D case,
for which $D_2=1.5$ is predicted, and $D_3=1.78\pm0.01$
(Charbonneau et al.~2001, McIntosh et al.~2002), $D_3=1.94\pm0.27, 
1.97\pm0.29$ (Aschwanden 2012a) for the 3D case, for which
$D_3=2.0$ is predicted. Thus, the mean value defined in Eq.~(9)
is a reasonably accurate prediction based on the standard (BTW) 
cellular automaton model.

This relationship (Eq.~9) allows also a scaling between the fractal
dimensions of the 2D and 3D Euclidean space,
\begin{equation}
	{D_3 \over D_2} \approx {(1+3) \over (1+2)} = {4 \over 3} \ .
\end{equation}

An extensive discussion of measuring the fractal geometry in SOC systems 
is given in Aschwanden (2011a, chapter 8) and McAteer (2013). 
Fractals are measurable from the spatial structure
of an avalanche at a given instant of time. Therefore, they enter 
the statistics of time-evolving SOC parameters, such as the observed 
flux per time unit, which is proportional to the number of 
instantaneously active nodes in a lattice-based SOC avalanche simulation.

\subsection{The Spatio-Temporal Evolution and Transport Process}

The next important step is to include time scales, which together
with the geometric scaling laws define the spatio-temporal evolution
of SOC events. We model a SOC event simply as an instability that
is triggered when a local threshold is exceeded. The universal 
behavior of any instability is an initial nonlinear growth phase
and a subsequent saturation phase. We model the saturation phase
with a diffusive function, as shown in Fig.~3 (upper panel),
\begin{equation}
	r(t) = \kappa (t - t_0)^{\beta/2} \ . 
\end{equation}
where $t_0$ is the onset time of the instability, $\kappa$ is the 
diffusion coefficient, and $\beta$ is the spreading exponent.
A value of $\beta \gapprox 0$ corresponds to logistic growth with
an upper limit of the spatial volume (Aschwanden 2011a, 2012b), 
$\beta \approx 0.5$ corresponds
to subdiffusion, $\beta=1$ to classical diffusion, $\beta \approx
1.5$ to hyper diffusion or L\'evy flight, and $\beta=2$ to linear
expansion.

The corresponding velocity $v(t)$ of an expanding SOC avalance is 
shown in Fig.~3 (second panel), which monotonically decreases with time 
and is obtained from the time derivative of $r(t)$ (Eq.~(11),
\begin{equation}
	v(t) = {d r(t) \over dt} = 
	{\kappa \beta \over 2}\ (t - t_0)^{\beta/2-1} \ .
\end{equation}

What spatio-temporal scaling law do we expect from this macroscopic
description of a SOC avalanche. A spatial scale $L$ could be defined
from the maximum size of the avalanche at the end time $T=(t-t_0)$, 
and thus we expect from Eq.~(11) the statistical spatio-temporal 
scaling law
\begin{equation}
	L \propto \kappa\ T^{\beta/2} \ .
\end{equation}
Substituting this scaling law $L(T)$ into the PFD of length scales
(Eq.~1), we expect a powerlaw distribution of time scales,
\begin{equation}
	N(T) dT = N(L[T]) {dL\over dT} dT
		= T^{-[1+(d-1)\beta/2]} = T^{-\alpha_T} \ .
\end{equation}
with the powerlaw slope of $\alpha_T=1+(d-1)\beta/2$,
which has a value of $\alpha_T =1 + \beta = 2.0$ for 3D-Euclidean space $(d=3)$ 
and classical diffusion ($\beta=1$). This powerlaw slope for avalanche
time scales is a prediction of universal validity, since it is only
based on the scale-free probability conjecture (Eq.~1), 
$N(L) \propto L^{-d}$, and the diffusive nature (or random-walk 
statistics) of the saturation phase.

The spatio-temporal scaling law (Eq.~13), based on random-walk 
or a diffusion process, is used here as a simple approximation in an 
empirical way. Diffusive transport has been applied to SOC theory
and SOC phenomena in a number of previous studies, e.g.,
by using the spreading exponents to determine the critical points 
of systems with multiple absorbing states (Grassberger and Delatorre 1979),
as a discretized diffusion process using the Langevin equation
(Wiesenfeld 1989; Zhang 1989; Foster et al.~1977; Medina et al. 1989),
in terms of classical (Lawrence 1991) and anomalous diffusion of 
magnetic flux events (Lawrence and Schrijver 1993),
in deriving spatio-temporal scaling laws with mean-field theory 
and branching theory (Vespignani and Zapperi 1998),
as a continuum limit of a fourth-order hyper-diffusive system
(Liu et al.~2001; Charbonneau et al.~2001), 
or in terms of a diffusion entropy description (Grigolini et al.~2002).

\subsection{	Energy and Flux Relationships		}

In numerical SOC simulations, such as in lattice-based cellular automaton 
models of the BTW type (Bak et al.~1987),
energy is dissipated in every node that exceeds a threshold temporarily,
and thus the energy that is dissipated during a SOC avalanche is
proportional to the total number of all active nodes, summed over space
at each instant of time. If we count these active nodes at a given
time interval, we have a quantity that is proportional to the instantaneous
energy dissipation rate, which has the unit of energy per time.
In the real world we observe a signal from a SOC avalanche in form of
an intensity flux $f(t)$ (e.g., seismic waves from earthquakes, hard X-ray 
flux from solar flares, or the amount of lost dollars per day in the 
stockmarket).
Let us assume that this intensity flux is proportional to the volume
of active nodes, which corresponds to the instantaneous fractal volume
$V_f(t)$ of a SOC avalanche in our spatio-temporal SOC model (Fig.~4),
also called {\sl fractal-diffusive (FD-SOC)} model (Aschwanden 2012a),
\begin{equation}
	f(t) \propto V_f(t) \propto r(t)^{D_d} \ ,
\end{equation}
which is shown in Fig.~3 (third panel) for $\beta=$0.1, 0.5, and 1.
The flux time profile $f(t)$ is expected to fluctuate substantially
in real data or in lattice simulations, because the fractal dimension 
can vary in the range of $D_{d,min} \approx 1$ and $D_{max}=d$, while
we use only the mean value $D_d=(D_{min}+D_{max})/2$ (Eq.~9) in our
macroscopic model. Occasionally, the instantaneous
fractal dimension may reach its maximum value, i.e., $D_d(t)\lapprox d$,
which defines an expected upper limit $f_{max}(t)$ of 
\begin{equation}
	f_{max}(t) \propto V(t) \propto r(t)^d \ .
\end{equation}

Integrating the time-dependent flux $f(t)$ over the time interval 
$[0,t]$ yields the total dissipated energy $e(t)$ up to time $t$
(using Eq.~11), 
\begin{equation}
	e(t) \propto \int_{0}^t V_f(t) dt = 
	             \int_{0}^t r^{D_d}(t) dt = 
	             \int_{0}^t \kappa^D_d (t-t_0)^{D_d \beta/2} dt = 
		     {\kappa^{D_d} \over D_d \beta/2+1} 
			(t-t_0)^{D_d \beta/2 + 1}\ ,
\end{equation}
which is a monotonically increasing quantity with time (Fig.~3, bottom 
panel).

From this time-dependent evolution of a SOC avalanche we can characterize
at the end time $t$ a time duration $T=(t-t_0)$, a spatial scale 
$L=r(t=t_0+T)$, an expected flux or energy dissipation rate $F=f(t=t_0+T)$, 
an expected peak flux or peak energy dissipation rate $P=f_{max}(t=t_0+T)$, 
and a dissipated energy $E=e(t=t_0+T)$, which is proportional to the
avalanche size $S$ in BTW models. Thus, we have the following scaling 
relations between the different SOC parameters and the length scale $L$
(using Eqs.~15-17),
\begin{equation}
	F \propto L^{D_d} \propto T^{D_d \beta/2} ,
\end{equation}
\begin{equation}
	P \propto L^d \propto T^{d \beta/2} \ ,
\end{equation}
\begin{equation}
	E \propto S \propto L^{D_d+2/\beta} \propto  T^{D_d \beta/2+1} \ .
\end{equation}
An alternative notation for the diffusive spreading exponent $\beta$  
used in literature is $D_T=2/\beta$, so that the spatio-temporal
scaling law (Eq.~13) reads as $T \propto L^{D_T}$ and the energy scaling
law (Eq.~20) as $E = S \propto V_f T \propto L^{D_d+D_T}$, which can be
expressed as $S \propto L^{D_S}$ with the exponent $D_S = D_3 + D_T$.  
Slight variations of this scaling law have been inferred from observations
in different wavelengths, such as $D_S = D_A/2 + D_T$ for magnetic
events (Eq.~18 in Uritsky et al.~2013), which seems to be equally
consistent with observations as our generalized (wavelength-independent)
FD-SOC model (see EIT and MDI events from Udritsky et al.~2013 in Table 1).

Finally we want to quantify the occurrence frequency distributions of the
the (smoothed) energy dissipation rate $N(F)$, the peak flux $N(P)$, 
and the dissipated energy $N(E)$, which all can readily be obtained by 
substituting the scaling laws (Eqs.~18-20) into the fundamental length scale
distribution (Eq.~1), yielding
\begin{equation}
        N(F) dF = N(L[F]) \left| {dL \over dF} \right| dF
        \propto F^{-[1+(d-1)/D_d]} \ dF \ ,
\end{equation}
\begin{equation}
        N(P) dP = N(/[P]) \left| {dL \over dP} \right| dP
        \propto P^{-[2-1/d]}
        \ dP \ ,
\end{equation}
\begin{equation}
        N(E) dE = N(L[E]) \left| {dL \over dE} \right| dE
        \propto E^{-[1+(d-1)/(D_d+2/\beta)]}
        \ dE \ .
\end{equation}
Thus this derivation from first principles predicts powerlaw functions for all
parameters $L$, $T$, $F$, $P$, $E$, and $S$, which are the hallmarks of SOC 
systems.  In summary, if we denote the occurrence frequency distributions
$N(x)$ of a parameter $x$ with a powerlaw distribution with power law index
$\alpha_x$,
\begin{equation}
        N(x) dx \propto x^{-\alpha_x} \ dx \ ,
\end{equation}
we have the following powerlaw coefficients $\alpha_x$ for the parameters
$x=L, T, F, P, E$, and $S$,
\begin{equation}
        \begin{array}{rl}
        \alpha_L &=  d \\
        \alpha_T &=  1+(d-1)\beta/2 \\
        \alpha_F &=  1+(d-1)/D_d \\
        \alpha_P &=  1+(d-1)/d \\
        \alpha_E = \alpha_S &=  1+(d-1)/(D_d+2/\beta)\\
        \end{array} \ .
\end{equation}
If we restrict to the case to 3D Euclidean space (d=3), as it is almost
always the case for real world data, the predicted powerlaw indexes are,
\begin{equation}
        \begin{array}{rl}
        \alpha_L &= 3 \\
        \alpha_T &= 1+\beta \\
        \alpha_F &= 1+2/D_3 \\
        \alpha_P &= 1+2/3 \\
        \alpha_E = \alpha_S &= 1+1/(D_3/2+1/\beta)\\
        \end{array} \ .
\end{equation}
Restricting to classical diffusion $(\beta=1)$ and an estimated mean
fractal dimension of $D_3 \approx (1+3)/2=2$ we have the following absolute
predictions
\begin{equation}
        \begin{array}{rl}
        \alpha_L &= 3 \\
        \alpha_T &= 2 \\
        \alpha_F &= 2 \\
        \alpha_P &= 5/3 \\
        \alpha_E = \alpha_S &= 3/2 \\
        \end{array} \ .
\end{equation}

\subsection{ Waiting Time Probabilities in the Fractal-Diffusive SOC Model }

The FD-SOC model
predicts a powerlaw distribution $N(T) \propto T^{-\alpha_T}$ of event
durations $T$ with a slope of $\alpha_T=[1+(d-1)\beta/2]$ (Eq.~25) that
derives directly from the scale-free probability conjecture $N(L) \propto
L^{-d}$ (Eq.~1) and the random walk (diffusive) transport
($L \propto T^{\beta/2}$; Eq.~13). For
classical diffusion ($\beta=1$) and space dimension $d=3$ the predicted
powerlaw is $\alpha_T=2$. From this time scale distribution we can also
predict the waiting time distribution with a simple probability argument.
If we define a waiting time $\Delta t$ as the time interval between 
the start time
of two subsequent events, so that no two events overlap with each other
temporally,  the waiting time cannot be shorter than the time
duration of the intervening event, i.e., $\Delta t_i \ge (t_{i+1}-t_i)$.
Let us consider the case of non-intermittent, contiguous flaring, but with 
no time overlap between subsequent events. In this case the waiting times
are identical with the event durations, and therefore their waiting
time distributions are equal also, reflecting the same statistical
probabilities,
\begin{equation}
        N(\Delta t) d\Delta t \propto N(T) dT
        \propto \Delta t^{-\alpha_{\Delta t}} d\Delta t \ ,
\end{equation}
with the powerlaw slope,
\begin{equation}
        \alpha_{\Delta t} = \alpha_T = 1 + (d-1)\beta/2 \ .
\end{equation}
This statistical argument is true regardless what the order of
subsequent event durations is, so it fulfills the Abelian property.
Now we relax the contiguity condition and subdivide the time series into
blocks with contiguous flaring (with intervals $\Delta t \approx T$), 
interrupted by arbitrarily long quiet periods $\Delta t = \Delta t^q$ 
when no events occur (Fig.~5). The distribution of quiet periods $\Delta t^q$ 
may be drawn from a random process, which has an exponential distribution
\begin{equation}
	N(\Delta t^q)  d \Delta t^q 
	\propto \exp{(-\Delta t^q / \Delta t^q_0)} \ d \Delta t^q  \ .
\end{equation}
If we define a maximum event duration $T_2$ and assume that this
is also approximately a lower limit for the quiet time intervals,
i.e., $\Delta t^q_{min} \approx T_2$, then we expect
a powerlaw distribution with a slope of $\alpha_{\Delta t} = \alpha_T$
for the range of waiting times that are shorter than the maximum
flare duration $\Delta t \lapprox T_2$, with an exponential cutoff 
at $\Delta t \gapprox T_2$. The contributions of waiting times 
from the subset of contiguous time blocks will still be identical, while
those time intervals from the intervening quiet periods add some longer
random waiting times. The predicted powerlaw slope of short waiting times 
($T_1 \lapprox \Delta t \lapprox T_2$) 
is then $\alpha_{\Delta t}=2.0$ for classical diffusion $\beta=1$ and space 
dimension $d=3$. Interestingly, this predicted slope is identical 
to that of nonstationary Poisson processes in the limit of intermittency 
(Aschwanden and McTiernan 2010). At the same time, this waiting time
model predicts also clustering of events during active periods,
and thus event statistics with memory and persistence, as it was
demonstrated recently for CME events using Weibull distributions
(Telloni et al.~2014).

\subsection{ Pulse Pile-Up Correction for Waiting Times	}	

We can define a mean waiting time $\langle \Delta t \rangle$
from the total duration of the observing period $T_{obs}$ and
the number of observed events
$n_{obs}$,
\begin{equation}
        \langle \Delta t \rangle = {T_{obs} \over n_{obs}} \ .
\end{equation}
From the distribution of event durations $T$, we have an inertial
range of time scales $[T_1, T_2]$, over which we observe a powerlaw
distribution, $N(T) \propto T^{-\alpha_T}$, with the corresponding
number of events $[N_1, N_2]$, so that we can define a nominal
powerlaw slope of $\alpha_T=\log(N_2/N_1)/\log(T_2/T_1)$.
If the mean waiting time of an observed time series becomes shorter
than the upper limit of time scales $T_2$ during very busy periods, 
we start to see
time-overlapping events, a situation we call {\sl ``event pile-up"}
or {\sl ``pulse pile-up''}. In such a case we expect that the waiting
time distribution starts to be modified, because the time durations
of the long events are underestimated (by some automated detection
algorithm), so that the nominal powerlaw slope that is expected with no
pulse pile-up, $\alpha_{\Delta t}=\log(N_2/N_1)/\log(T_2/T_1)$,
has to be modified by replacing the lower time scale $T_1$ with the
mean waiting time $\langle \Delta t \rangle$,
\begin{equation}
        \alpha_{\Delta t}^{pileup} \ = \ \alpha_{\Delta t}
        \times \left\{
        \begin{array}{ll}
                1 & {\rm for}\ \langle \Delta t \rangle \ > \ T_2 \\
                \log(T_2)/\log{\langle \Delta t \rangle}
                  & {\rm for}\ \langle \Delta t \rangle \ \le \ T_2 \\
        \end{array}
        \right.
\end{equation}
As a consequence, the measurements of event durations must suffer from 
the same pile-up effect, and a similar correction is expected
for the time scale distribution $N(T)$,
\begin{equation}
        \alpha_T^{pileup} \ = \ \alpha_T
        \times \left\{
        \begin{array}{ll}
                1 & {\rm for}\ \langle \Delta t \rangle \ > \ T_2 \\
                \log(T_2)/\log{\langle \Delta t \rangle}
                  & {\rm for}\ \langle \Delta t \rangle \ \le \ T_2 \\
        \end{array}
        \right.
\end{equation}
Thus the predicted waiting time distribution has a slope of
$\alpha_T=2$ in the slowly-driven limit, but can be
steeper in the strongly-driven limit. For instance, the
waiting time distributions of solar flares correspond to the
slowly-driven limit during the minima of the solar 11-year cycle,
while the powerlaw slopes indeed steepen during the maxima of the
solar cycle (Aschwanden and Freeland 2012), when the flare density 
becomes so high that the slowly-driven limit, and thus the separation 
of time scales, is violated.

\subsection{Physical Scaling Laws}

Our fractal-diffusive SOC model developed so far has universal validity
because it is entirely derived from statistical probabilities and
fractal-diffusive transport. The predicted scaling laws and occurrence
frequency distributions derived above do not depend on any specific
physical parameter of a SOC phenomenon. Using real-world observations,
however, some physical scaling laws are involved between the observables
and the spatio-temporal parameters used so far. For instance, the strength
of an earthquake is measured in magnitudes of the Gutenberg-Richter scale
(Richter 1958), which may be related to the observed earthquake rupture 
area by some mechanical scaling law that determines the statistics
(Main and Burton 1984). For solar flares, the observed fluxes in soft or
hard X-rays are related to the physical parameters of electron temperatures,
densities, and pressures of heated plasma, as it can be derived for the
equilibrium point between heating and cooling (e.g., Rosner et al.~1978).
Other scaling laws used in solar physics include, for instance, relationships
between magnetic energies and the reduced MHD equations (Longcope and Sudan
1992), or the magnetic reconnection geometry (Craig 2001), or between the
heating rate and the magnetic field strength (Schrijver et al.~2004).
Such physical scaling laws allow us to derive the powerlaw slope of the 
frequency distribution of both the observables and the physical parameters, 
which is examined elsewhere (e.g., Aschwanden et al.~2013). 

The predicted frequency distributions for energies and fluxes
derived in Section 2.5, are strictly only valid for systems where the
assumption of proportionality between the flux and the instantaneous 
fractal volume
is fulfilled, i.e., $F(t) \propto V_f(t) \propto r(t)^{D_d}$ (Eq.~15),
because the scaling of observables depends then on geometric parameters
only, which can be derived entirely from statistical probabilities,
in terms of the scale-free probability conjecture (Eq.~1).

Without specializing on a particular physical mechanism of a given
SOC system, we can give some general rules how to derive the powerlaw
function of physical parameters. The simplest situation is a 
2-parameter correlation or scaling law, where a physical parameter $x$
is related to the geometric length scale by a powerlaw function with 
index $\gamma$,
\begin{equation}
	L \propto x^\gamma \ .
\end{equation}
Inserting this scaling law into the fundamental length scale distribution
$N(L) \propto L^{-d}$ (Eq.~1) and using the derivative $dL/dx = x^{\gamma-1}$
yields then directly the occurrence frequency distribution $N(x)$,
\begin{equation}
	N(x) dx = N(L[x]) {dL \over dx} dx = x^{-[1+(d-1)\gamma]} dx \ .
\end{equation}
Also common is a 3-parameter correlation or scaling law,
such as in terms of two physical parameters $x$ and $y$ and the length
scale $L$, i.e.,
\begin{equation}
	L \propto x^\gamma y^\delta \ ,
\end{equation}
in terms of powerlaw functions with exponents $\gamma$ and $\delta$.
The probability distribution for one of the physical parameters, say $y$,
can then be written as, 
\begin{equation}
	N(y) dy = \int N(L, x[L,y]) dL \ dy
		= \int N(L) N(x[L,y]) dL \ dy
		\propto y^{a_x \delta / \gamma} \ dy \ ,
\end{equation}
after the integration over the variable $L$ is carried out.
Thus the resulting distribution $N(y) dy \propto y^{-\alpha_y} dy$ has a
powerlaw slope of $\alpha_y = \alpha_x \delta / \gamma$. 
The powerlaw solution is strictly valid only for complete sampling
of the parameters, which in reality is often not possible due to
limited statistics, instrumental sensitivity limits, and data noise. 
This leads to truncation effects and finite-size effects, which can
be simulated with Monte-Carlo simulations or analytically calculated
(see Appendix in Aschwanden et al.~2013 for examples).

\subsection{Instrument-Dependent Size Distributions}

Besides physical scaling laws that are specific to a particular
physical mechanism of a SOC system, there are also instrument-dependent
scaling laws that are not universal and depend on the specific instrument
used in an observation of SOC phenomena. If there is a nonlinear scaling
between the observable and the geometric volume of a SOC avalanche,
we cannot expect to measure the same powerlaw slope of an observable
with different instruments. In order to make observed frequency
distributions obtained with different instruments compatible, it is
often advisable to reduce the observable parameters to physical
parameters using a well-established instrument calibration.
For astrophysical observations in soft X-rays and EUV, for instance,
an instrument-independent physical quantity is the differential
emission measure distribution, which can be inverted from observed
fluxes in different wavelengths (e.g., Aschwanden et al. 2013).

\section{RELATIONSHIP TO THEORETICAL MICROSCOPIC SOC MODELS}

After we have described a general macroscopic model of a SOC system
that predicts the occurrence frequency distributions of spatial,
temporal, and volume-related observables, such as the flux and energy,
we turn now to theoretical and numerical SOC models and discuss 
whether our macroscopic model meets the basic definitions of a
SOC system. A comprehensive review of theoretical and numerical
SOC models is given in the textbook by Pruessner (2012). While
a strict definition of SOC systems is still not well-established,
we will use here the working definition given in the Introduction:
{\sl SOC is a critical state of a nonlinear energy dissipation system that 
is slowly and continuously driven towards a critical value of a system-wide
instability threshold, producing scale-free, fractal-diffusive, and 
intermittent avalanches with powerlaw-like size distributions.} 
The property of self-tuning
to criticality is warranted by system-inherent physical conditions 
that define a system-wide instability threshold. This system-inherent
physical condition is often given by the equilibrium solution between
two competing forces. For instance, the angle of repose in a sandpile 
is self-tuning to a system-wide critical value, corresponding to an
equilibrium point between the gravity force and the static friction 
force. In the Ising model (Ising 1925), a phase transition occurs
at a critical point between an ordered and a disordered magnetic 
spin state, but the tuning to the critical point is not 
self-organized.  In the following we discuss how the macroscopic
SOC model (Section 2) relates to the microscopic (mathematical
and numerical) SOC models, regarding powerlw-scaling (Section 3.1),
spatio-temporal correlations (Section 3.2), separation of time scales 
and intermittency (Section 3.3), and self-tuning and critical threshold 
(Section 3.4).

\subsection{Powerlaw Scaling}

The original Bak-Tang-Wiesenfeld (BTW) model revealed the
generic scale invariance of simulated or observed SOC parameters,
which ideally exhibits powerlaw functions for the occurrence frequency 
distributions, possibly related to the 1/f-noise of power spectra
(Bak et al.~1987). 
The property of a powerlaw shape became the hallmark
of SOC phenomena, but it was recognized that this is a necessary
but not a satisfactory condition, since other phenomena (such as
turbulence or percolation) produce powerlaws also. 

Our fractal-diffusive SOC model (FD-SOC) derives the probability
distribution functions (PFD) based on a statistical probability
argument, which leads to a powerlaw function of spatial and geometric
scales. The additional assumption of fractal-diffusive transport
leads to a powerlaw function of temporal scales. Further we
define the size of an avalanche from the time-integrated fractal 
volume that participates in an avalanche, and consequently we obtain
also powerlaw distributions for the size or total dissipated energy
of avalanches.
Since all these assumptions are of statistical nature and do not 
depend on any physical parameters of a SOC system, the predictions 
of the PDFs of spatial, temporal, and energy SOC parameters have 
universal applicability, irregardless of the physical process that
is involved in the nonlinear energy dissipation process. 
The prediction of a pure powerlaw function for the size distributions
at all scales is also called {\sl universality} in theoretical
SOC models (e.g. Sethna et al.~2001), and is fulfilled in the 
macroscopic description of our FD-SOC model by design (as a 
consequence of the scale-free probability conjecture; Eq.~1). 
However, we should be aware that this 
simple FD-SOC model provides only a first-order prediction, while 
additional effects (such as truncation, incomplete sampling, or 
finite-size effects) may modify the observed size distributions
into broken powerlaws, double powerlaws, or other powerlaw-like
distribution functions. However, similar effects occur also in
cellular automaton simulations.

\subsection{Spatio-Temporal Correlations}

SOC systems are expected to exhibit spatio-temporal correlations
(Jensen 1998) of a SOC state variable ${\bf B({\bf r},t)}$, 
\begin{equation}
	C({\bf r}, t) = 
	<B({\bf r}_0, t_0)
	 B({\bf r}_0+{\bf r}, t_0+t) >
	- < {\bf B}({\bf r}_0, t_0) >^2 \ .
\end{equation}
Such correlations are absent in systems with random noise. 
In our FD-SOC model, however, the random structure of the
background in a state near criticality is episodically disturbed 
by an avalanche event, which carves out a ``hole'' with a size $L$
during a time scale $T$, which represents a major disturbance in form
of a spatially and temporally coherent structure, which can be restored
to the critical state only gradually, for slowly-driven SOC systems.
Naturally, large avalanches leave their footprints behind and
produce spatio-temporal correlations during the local restoration time. 
The correlation is best for large avalanches with similar 
shapes. The time profiles of avalanches in our FD-SOC system are
self-similar to some extent, since they are characterized 
by a common fractal dimension $D_d$ (Eq.~15), diffusion constant 
$\kappa$, and diffusive spreading exponent $\beta$ (Eq.~11). We visualize 
the spatial correlations with a cartoon in Fig.~6, which shows 
coherent disturbances as deviations from the critical state in 
large avalanches occurring in sandpiles and in solar flares. 

In our two-component model of waiting times (Section 2.6), 
an observed time series consist of quiet intervals $\Delta t^q > T_2$
with no avalanching (which have a random distribution), and active
intervals with contiguous flaring (which have a powerlaw distribution
like the event durations $N(\Delta t) = N(T)$). This dual behavior
is also called {\sl intermittency}, and has the consequence that
the combined waiting time distribution has both a powerlaw range
($T_1 \le \Delta t \le T_2$) and an exponential cutoff ($\Delta t \ge T_2$).
Consequently, we expect spatio-temporal correlations (Eq.~38) 
during the intermittently active periods only, while they are expected
to be absent during the quiet time intervals. Avalanching during
active periods is also expected to exhibit persistence and memory,
while no memory is expected during quiet time intervals. This
property seems to be more consistent with observations (e.g.,
Telloni et al.~2014), but is different from the pure random
(Poisson) statistics of the original BTW model, but reconciles
related debates about the functional shape of the waiting
time distributions (e.g., Boffetta et al.~1999; Lepreti et al.~2001).

\subsection{Separation of Time Scales and Intermittency}

Classical SOC systems operate in the limit of slow driving, which
implies a separation between the duration of an avalanche and the
waiting time interval between two subsequent avalanches.
Numerically, the separation of time scales is simply realized by
allowing only one single disturbance of a SOC system at a time,
which triggers an avalanche (with duration $T$) or not, while the  
next disturbance is not initiated after a waiting time $\Delta t > T$, 
in the case of an avalanche. 

In our FD-SOC model, the energy dissipation rate $(de/dt)$ is
monotonically growing after a triggering disturbance, which 
exceeds the system-wide threshold value $(de/dt)_{crit}$ until
the spatial diffusion stops after time $T$, due to a lack of
unstable nodes among the next-neighbors of an instantaneous
avalanche area or volume. Therefore, the energy dissipation rate
during an avalanche exceeds the threshold value during the entire
duration of an avalanche. Energy conservation between the 
slowly-driven energy input rate and the intermittent avalanching
output rate can therefore only be obtained with sufficiently long
waiting times $\Delta t$ during which the energy loss of an
avalanche is restored. This requires a balance of the long-term
averages of the energy input and output rates, i.e.,
\begin{equation}
	\langle (de/dt)_{in} \rangle  \langle \Delta t \rangle 
	\approx \langle (de/dt)_{out} \rangle \langle T \rangle \ .
\end{equation}
Since $\langle (de/dt)_{in} \rangle \le (de/dt)_{crit} \ll 
\langle (de/dt)_{out} \rangle $, 
it follows that $ \langle \Delta t \rangle \gg \langle T \rangle $, 
which warrants a separation of the time scales, i.e., the
waiting time $\Delta t$ and the avalanche duration $T$.

The resulting time profile of the energy dissipation rate $(de/dt)$
of a SOC system is then necessarily highly intermittent due to the
long waiting times inbetween subsequent avalanches. In addition,
the time profile is strongly fluctuating during an avalanche,
according to $f(t) \propto r(t)^{D_d}$ (Eq.~15), since the fractal
dimension $D_d(t)$ can fluctuate in the entire range between the 
minimum and maximum
value as a function of time, i.e., $1 \lapprox D_d(t) \le d$. 
However, for the scaling laws in the FD-SOC model (Eqs.~18-20),
we can replace the fluctuating value of $D_d(t)$ with a constant
mean value $\langle D_d(t) \rangle = (1+d)/2$ and obtain the same
size distributions. 

\subsection{Self-Organization and Criticality}

How does our fractal-diffusive SOC model reinforce self-organized
criticality? In classical SOC models, criticality is obtained
by a slowly-driven input of energy which restores the energy losses
of avalanches until the system-wide critical threshold is reached
(more or less) and new avalanches can be triggered by a local excess
of the critical threshold. In our FD-SOC model, the time evolution
of an avalanche has a generic shape that is given by fractal-diffusive 
transport, while the energy balance between energy input 
(disturbances) and output (avalanches) is not explicitly reinforced,
unlike cellular automaton models which iterate a mathematical redistribution
rule to drive the dynamics of a SOC system and are designed to conserve
energy. Instead, self-organization of the FD-SOC model is constrained by 
statistical probability only, which does not need to be self-tuning to produce 
a particular functional form of a size distribution, because there is only one 
statistical distribution with maximum likelyhood, which is a powerlaw
distribution function of spatial scales (according to our scale-free 
probability conjecture). So, we can say that the FD-SOC model gravitates around
the statistically most likely state, like entropy in self-contained
statistical systems without external influence. This may be a more
general definition of self-organized criticality then originally
proposed by Per Bak and coworkers, but explains the concept of 
self-organization by the most general principle of maximum statistical 
likelyhood. This should not surprise us, since the entire evolution 
of our universe followed maximum statistical likelihood, from the 
initial big bang expansion all the way to the bio-chemical evolution 
of life, forming complexity out of simple structures based on 
processes that are driven by statistical likelyhood (e.g., Mendel's
law in genetics). 

\section{ASTROPHYSICAL APPLICATIONS}

In this section we examine frequency distributions observed in
various realms of astrophysics and discuss the application of
the fractal-diffusive SOC model in a few selected datasets 
with large statistics. Some preliminary discussion of such 
astrophysical objects can also be found in Aschwanden (2011a; 
chapters 7 and 8) and in Aschwanden (2013; chapter 13). An 
overview of astrophysical phenomena with observed powerlaw 
indices of size distributions is given in Table 1.

\subsection{Lunar Craters}

If we mount a large container with a gel-like surface below a circular 
plate that holds Per Bak's sandpile, we would record impact craters 
from each sandpile avalanche in the viscous gel and could infer 
the avalanche sizes from the diameters of the impact craters
(see experimental setup of sandpile experiment conducted by 
Held et al.~1990).
Similarly, the Moon was targeted by many impacting meteors and
meteorites, especially during an intense bombardment in the
final sweep-up of debris at the end of the formation of the solar
system between 4.6 and 4.0 billion years ago (e.g., Neukum et al.~2001). 
The sizes of lunar
craters were measured with the first lunar orbiters (Ranger 7, 8, 9)
in the early 1960's, and a cumulative powerlaw distribution with
sizes in the range of $L \approx 10^0 - 10^{4.5}$ cm was found, with
a powerlaw slope of $\alpha^{cum}_L \approx 2.0$ for the 
cumulative distribution (Cross 1966), which corresponds to a value of 
$\alpha_L=\alpha^{cum}_L+1=3.0$ for the differential
size distribution. This quite accurate result (for a size
distribution covering a range of over 4 orders of magnitude) corresponds
exactly to our prediction of the scale-free probability conjecture,
$N(L) \propto L^{-3}$ (Eq.~1). A similar value of $\alpha_L=2.75$
was found for the size distribution of meteorites and space debris
from man-made rockets and satellites (Fig.~3.11 in Sornette 2004).
The formation of the sizes of meteors 
and meteorites may have been controlled by a nonlinear process that
includes a combination of
self-gravity, gravitational disturbances, collisions, depletions, 
fragmentation, and captures of incoming new bodies in the solar system
(e.g., Ivanov 2001). 
The Moon acts as a target that records the sizes of impacting
meteorites that were produced by a SOC process, similar to the
gel-filled plate under Bak's sandpile.

\subsection{Asteroid Belt}

The origin of the asteroid main belt is believed to be associated with
a time period of intense collisional evolution shortly after the 
formation of the planets (e.g., Botke et al.~2005). The asteroids
are a leftover of the planetesimals that were either too small
to form a planet by self-gravitation, or they orbited in an unstable
region of the solar system that constantly got disturbed by the 
largest planets Jupiter and Saturn.

In Table 1 we compile some values of measured size distributions
of asteroids, given as powerlaw slopes $\alpha_L$ of the differential 
size distributions (related to the slope $\alpha^{cum}_L$ of the cumulative 
size distribution by $\alpha_L=\alpha^{cum}_L+1$, which includes values in
the range of $\alpha_L \approx 2.3-4.0$, obtained from the 
Palomar Leiden Survey (Van Houten et al.~1970), the Spacewatch
Surveys (Jedicke and Metcalfe 1998), the Sloan Digital Sky Survey
(Ivezic et al.~2001), and the Subaru Main-Belt Asteroid Survey
(Yoshida et al.~2003; Yoshida and Nakamura 2007). These values
of the powerlaw slopes agree within $\approx 25\%$ with our
theoretical prediction of $\alpha_L=3.0$, but the statistical
range of sizes covers less than two decades, and thus incomplete
sampling of small sizes is likely to limit the accuracy.

\subsection{Saturn Ring}

The Saturn ring extends over a range of 7,000-80,000 km above Saturn's
equator and has a mass of $3 \times 10^{19}$ kg, consisting of myriads 
of small particles with sizes in the range from 1 mm to 20 m
(Zebker et al.~1985; French and Nicholson 2000). The particle size
distribution was measured in eight different ring regions with
{\sl Voyager I} radio occultation measurements (Zebker et al.~1985). 
These size distributions were found to have slightly different
powerlaw slopes in each ring zone, with values of $\alpha_L
=2.74 - 3.03$ for ring A, $\alpha_L=2.79$ for the Cassini division,
and $\alpha_L=3.05-3.22$ for ring C (Zebker et al.~1985). Averaging
the values from all eight zones we find $\alpha_L=2.89\pm0.16$,
which is remarkably close to the prediction $\alpha_L=3.0$ of the
scale-free probability conjecture (Eq.~1). Thus, the fragmentation
of Saturn ring particles is consistent with the statistics of
SOC avalanches, and the process of collisional fragmentation
driven by celestial mechanics can be considered as a self-organizing
system that is constantly driven towards the collisional instability 
threshold. An instability occurs by a collision of particles. If the
system has a too low density, no collisions occur and the system
is subcritical, while a too high density of particles would result
into an excessive collision rate that would destroy the structure
of the Saturn ring. Hence, the long-lived Saturn ring can be
considered as a SOC system that self-tunes to a critical collisional
limit that maintains its shape and conserves its (kinetic) energy, 
similar to Bak's SOC sandpile that maintains its slope and conserves
the potential energy.

\subsection{Magnetosphere}

The Earth's magnetosphere displays a number of phenomena that have
been associated with SOC models (Table 1), such as active and quiet
substorms and auroral events (Lui et al.~2000; 
Uritsky et al.~2001, 2002, 2006; Kozelov et al.~2004; 
Klimas et al. 2010), substorm flow bursts 
(Angelopoulos et al. ~1999), auroral electron (AE-index) bursts
(Takalo 1993; Takalo et al.~1999), upper auroral (AU-index) bursts
(Freeman et al.~2000; Chapman and Watkins 2001), or outer radiation 
belt electron events (Crosby et al.~2005). The powerlaw indexes
of observed size distributions of these phenomena are listed in
Table 1. 

Accurate measurements, using the same definition of
time-integrated avalanche sizes as in the BTW model (Bak et al.~1987;
Charbonneau et al.~2001) and in this paper, were carried out
for auroral events in UV by Uritsky et al.~(2002), and in visible
light by Kozelov et al.~(2004), yielding size distribution with
powerlaw slopes of $\alpha_A \approx 1.7-2.0$, $\alpha_T \approx
2.0-2.5$, $\alpha_P \approx 1.66-2.0$, and $\alpha_E=1.4-1.7$,
which agree well with the predictions of the FD-SOC model
($\alpha_A = 2.0$, $\alpha_T = 2.0$, $\alpha_P \approx 1.67$, 
$\alpha_E=1.5$) (Table 1). The earlier reported lower values 
for the powerlaw slopes of auroral fluences (Lui et al.~2000)
are incompatible with recent observational results as well as
with the FD-SOC model, because the auroral sizes were measured 
from snapshots taken in regular time intervals, rather than 
measured individually for each avalanche event (Udritsky et al.~2002). 
This case with contradicting statistical results measured 
from the same data is an example of a validation test with 
the FD-SOC model.

The number of electrons in the outer radiation belt 
(at $4-8$ L-shell distances) is modulated by the solar wind,
exhibiting size distributions of electron peak fluxes with powerlaw
slopes of $\alpha_P \approx 1.5-2.1$ (Crosby et al.~2005).
The variation of the powerlaw slope is mostly attributed to
variations of the orbits of the microsatellites (STRV-1a and 1b)
that record the electron bursts at different intersections of
the radiation belt with the orbits. Nevertheless, the mean
value averaged over different years and L-shell distances,
$\alpha_P = 1.7\pm0.2$, is quite consistent with the theoretical
prediction $\alpha_P=1.67$ of the FD-SOC model. The radiation belt
can be considered as a SOC system, where the input is driven by 
solar wind electrons, which become trapped in the outer radiation belt, 
while magnetic variations modulate the untrapping of electrons
by a self-organizing loss-cone angle, producing avalanches of 
electrons bursts. 

\subsection{Solar Flares}

Solar flares have been interpreted as a SOC phenomenon since 1991
(Lu and Hamilton 1991) and numerous studies have been performed
to establish the size distributions of various solar flare parameters
measured in hard X-rays (HXR), soft X-rays (SXR), extreme ultraviolet
(EUV), and radio wavelengths. A representative selection of powerlaw 
slopes from size distributions of solar flare length scales ($\alpha_L$), 
flare areas ($\alpha_A$), time durations ($\alpha_T$), peak fluxes
($\alpha_P$), and fluences or energies ($\alpha_E$) is given in Table 1
(see references in footnote of Table 1).
We note that most of the powerlaw slopes measured in HXR, SXR and
EUV agree well with the theoretical predictions of our FD-SOC model, 
i.e., $\alpha_L=3.0$, $\alpha_A=2.0$, $\alpha_T=2.0$, $\alpha_P=1.67$,
and $\alpha_E=1.50$, say typically within 5\% to 10\%. The remaining
differences can be attributed to the different instrumental bias
and the different analysis methods (threshold definition, preflare
background subtraction, temperature bias) of the observations.
Also the peak fluxes observed in radio wavelengths are commensurable 
with the predictions for incoherent emission mechanisms, such as
gyrosynchrotron emission in microwave bursts. 
Only the solar energetic particles appear to have 
a flatter distribution than predicted, which has been interpreted in
terms of a selection bias for large events (Cliver et al.~2012), or
alternatively in terms of the geometric dimensionality of the SOC
system (Kahler 2013). In summary, except for the SEP events,
solar flares observed in almost all wavelengths are in agreement
with the FD-SOC model and provide the strongest support for SOC
models among all astrophysical phenomena.

What are the physical mechanisms in a SOC system that produce solar 
flares. The solar corona is considered to be a multi-component SOC 
system, where each active region or quiet Sun region represents a
different SOC sandpile, with its own spatial (finite-size) boundary,
lifetime, and flaring rate. Interestingly, the statistics of a single
SOC system (one active region) seems not to be significantly different
from the statistics of an ensemble of SOC systems (in the entire corona),
except for a different largest-event cutoff (Kucera et al.~1997).
The energy input comes ultimately from
build-up of nonpotential magnetic fields (with electric currents)
that is driven by subphotospheric magneto-convection and magnetic
flux emergence. The coronal
SOC system is slowly driven by continuous emergence of magnetic flux,
braiding, and stressing of the magnetic field. Once a local threshold
for instability is exceeded (kink instability, torus instability,
tearing mode instability, etc.), an avalanche of magnetic energy 
dissipation is triggered that ends in a fractal-diffusive phase.
The dissipated energy can be converted into thermal energy of heated
plasma (visible in soft X-rays and EUV), and into kinetic energy of
accelerated particles (detectable in hard X-rays and in gyrosynchrotron
emission in radio wavelengths). The fact that we measure similar
powerlaw slopes in all wavelengths (HXR, SXR, EUV, radio) implies
that all converted energies are approximately proportional to the
emitting volume, i.e., $E \propto V_f T$. Only solar energetic particles
(SEP) events and coherent emission in radio wavelengths show a much 
flatter powerlaw slope, which indicates a nonlinear scaling law 
$E \propto (V_f T)^\gamma$ or a selection bias for large events. 

\subsection{Stellar Flares}

Stellar flares have been observed in small numbers during a few hours
with the {\sl Hubble Space Telescope (HST)} (Robinson et al.~1999),
the {\sl Extreme Ultraviolet Explorer (EUVE)} (Audard et al.~2000; 
Kashyap et al.~2002; G\"udel et al.~2003; Arzner and G\"udel 2004;
Arzner et al.~2007), and the {\sl X-ray Multi-Mirror Mission (XMM) - Newton}
(Stelzer et al.~2007), which produced size distributions of flare
energies (time-integrated EUV fluxes) with powerlaw slopes in a
range of $\alpha_E = 2.17 \pm 0.25$ (Table 1). These values are significantly
steeper than derived for solar flare energies ($\alpha_E \approx
1.5-1.6$), but are expected for small samples near the exponential 
fall-off at the upper end of the size distribution (Aschwanden 2011a).
In addition, since plasma cooling extends the soft X-ray and EUV flux
beyond the time interval of energy release, the fluence of the largest
flares may be over-estimated for the largest solar and most stellar 
flare events.

Much larger statistics of stellar flares became recently available
from the {\sl Kepler} mission: 373 flaring stars were identified
in a search for white-light flares on $\approx 23,000$ cool dwarfs
in the {\sl Kepler} Quarter 1 long cadence data (Walkowicz et 
al.~2011; Maehara et al.~2012); a total of 1547 superflares (several
orders of magnitude larger than solar flares) were detected on
279 G-type (solar-like) stars (Notsu et al.~2013; Shibayama
et al.~2013). The flare energies were estimated from the 
time-integrated bolometric luminosity in visible light. 
Similar energy size distributions were found as in earlier smaller
samples (with EUVE), with powerlaw slopes of $\alpha_E=2.0\pm0.2$
for flares on all G-type stars, and $\alpha_E=2.3\pm0.3$ for
flares on slowly rotating G-type stars (Maehara et al.~2012;
Shibayama et al.~2013). We show the size distribution for the
total sample of 1538 stellar flares in Fig.~7 (middle panel), 
which has a powerlaw slope of $\alpha_E=2.04\pm0.13$. From
Kretzschmar (2011, Table 1 therein) we derive a scaling law
between the bolometric fluence (total solar irradiance) 
$E_t$ (which is equivalent to the bolometric energy $E_b$) 
and the GOES 1-8 \ang\ peak flux $P_x$ (Fig.~7, top panel),
\begin{equation}
	E_b \propto E_t \propto P_x^{(0.78\pm0.13)} \ .
\end{equation}
Using this scaling law we can derive the distribution of GOES
peak fluxes of the stellar flares,
\begin{equation}
	N(P_x) d P_x \propto N(E_b[P_x]) 
	{dE_b \over d P_x}\ d P_x   
	\propto P_x^{-1.81\pm0.12}\ d P_x \ ,
\end{equation}
which is consistent with the size distribution of GOES fluxes 
directly obtained by applying the scaling law of Kretzschmar (2011)
given in Eq.~(40), with a powerlaw slope of $\alpha_P=1.88\pm0.09$
(Fig.~7, bottom). Interestingly, the so obtained peak flux 
$\alpha_P \approx 1.88$ agrees better with the theoretical 
prediction $\alpha_P = 1.67$ of our FD-SOC model, than the
bolometric fluence. This may indicate that the bolometric fluence is not
an accurate proxy of the flare energy or flare volume, possibly
due to a nonlinear scaling of the bolometric fluence with flare
energies. 

\subsection{Pulsars}

Pulsars exhibit intermittent irregular radio pulses, besides
the regular periodic pulses that are synchronized with their
rotation period. The irregular pulses indicate some glitches
in the positive spin-ups of the neutron star, possibly caused
by sporadic unpinning of vortices that transfer momentum to 
the crust (Warzawski and Melatos 2008), which was interpreted
as a SOC system (Young and Kenny 1996). A size distribution
of the radio fluxes from the Crab pulsar exhibited a powerlaw
distribution with slopes in the range of $\alpha_P=3.06-3.50$
(Argyle and Gower 1972; Lundgren et al.~1995). A similar value
of $\alpha_P=2.8\pm0.1$ was found for PSR B1937+21 (Cognard
et al.~1996). Statistical measurements of the size distribution 
of pulsar glitches obtained from about a dozen of other pulsars 
yielded a large scatter of values in the range of 
$\alpha_P=-0.13,...,2.4$ (Melatos et al.~2008). The reasons for 
these inconsistent values may be rooted in the small-number 
statistics ($N=6,...,30$) and methodology (rank-order plots).
If the more reliable values from the crab pulsar hold up
($\alpha_P \approx 3.0$), which are typical for size distribution
of length scales ($\alpha_L=3.0$), physical 
models that predict a proportionality between peak fluxes $P$ and
length scales $L$ should be considered. 

\subsection{Soft Gamma-Ray Repeaters}

Soft gamma-ray repeaters, detected at energies of $>25$ keV with
the {\sl Compton Gamma Ray Observatory (CGRO)} and the {\sl Rossi
X-ray Timing Explorer}, exhibited size distributions with fluences
in the range of $\alpha_E=1.43-1.76$ (Gogus et al.~1999, 2000),
which is quite consistent with the values measured from solar flares
at the same energies and predicted by our FD-SOC model ($\alpha_E=1.5$).
However, the physical mechanisms of soft gamma-ray repeaters are
entirely different from solar flares, believed to originate from
slowly rotating, extremely magnetized neutron stars that are
located in supernova remnants (Kouveliotou et al.~1998, 1999), 
where neutron star crust fractures occur, driven by the stress of an 
evolving, ultrastrong magnetic field in the order of 
$B \gapprox 10^{14}$ G (Thompson and Duncan 1996). The fact
that solar flares and soft gamma-ray repeaters exhibit the same energy
size distribution, although the underlying physical processes are
entirely different, supports the universal applicabilitity of our 
FD-SOC model.

\subsection{Black Hole Objects}

Cygnus X-1, the first galactic X-ray source that has been identified
as a black-hole candidate, emits hard X-ray pulses with a time variability
down to 1 ms, which is attributed to bremsstrahlung X-ray pulses from
mass infalling towards the black hole and the resulting turbulence in
the accretion disk. Observations with {\sl Ginga} and {\sl Chandra}
exhibit complex 1/f noise spectra and size distributions of peak fluxes
with very steep powerlaw slopes of $\alpha_P \approx 7.1$ (Negoro
et al.~1995; Mineshige and Negoro 1999), which have been interpreted
in terms of SOC models applied to accretion disks (Takeuchi et al.~1995; 
Mineshige and Negoro 1999). Such steep values of the powerlaw slope
of peak fluxes are difficult to understand in terms of our standard
FD-SOC model, which predicts $\alpha_P=1.67$. They exclude a linear
scaling between the peak flux $P$ and the emitting volume $V$ covered
by an X-ray pulse. Such a steep slope can only be produced by an
extremely weak dependence of the X-ray peak flux $P$ on the avalanche
volume $V$, requiring a quenching mechanism that limits every
fluctuation to almost the same level. The cellular automaton model
of Mineshige and Negoro (1999), which can produce powerlaw size
distributions with such steep slopes of $\alpha_P \approx 7$, indeed
prescribes a non-random distribution of time scales for large pulses
(shots), where the occurrence of large pulses is suppressed for a
certain period after each large pulse.

\subsection{Blazars}

Blazars (BL Lacertae objects) are high-polarization quasars and
optically violent variable stars, which exhibit a high degree of
fluctuation in radio and X-ray emission due to their particular
orientation with the jet axis almost coaligned with our line-of-sight.
Light curves from GX 0109+224 were analyzed and found to exhibit
a 1/f noise spectrum, i.e., $P(\nu) \propto \nu^{-p}$ with $p=1.57-2.05$,
and a size distribution of peak fluxes with a powerlaw slope of
$\alpha_P=1.55$, and have been interpreted in terms of a SOC model
(Ciprini et al.~2003). This value is quite consistent with the
prediction of our standard FD-SOC model ($\alpha_P=1.67$), which
suggests that the peak flux $P$ emitted (in optical and radio
wavelengths) is proportional to the emitting volume $V$.  
The agreement between observations and the theoretical prediction 
supports the universal applicability of the FD-SOC model. 

\subsection{Cosmic Rays}

Cosmic rays are high-energetic particles that propagate through a
large part of our universe and are accelerated by galactic and 
extragalactic magnetic fields. Cosmic ray energy spectra span 
over a huge range of $E=10^9-10^{21}$ eV, where the lower limit
of $\approx 1$ GeV corresponds to the largest energies that can
be accelerated in solar flares and coronal mass ejections.
This cosmic ray energy spectrum exhibits an approximate powerlaw 
function with a mean slope of $\alpha_E \approx 3.0$ (Fig.~8, bottom right).
A more detailed inspection reveals actually a broken powerlaw
with a slope of $\alpha_{E1}\approx 2.7$ below the knee at
$E_{knee}\approx 10^{16}$ eV, and a slope of $\alpha_{E2}\approx
3.3$ above the knee. The two energy regimes are associated with
the particle origin in galactic space ($E \lapprox E_{knee}$) and
extragalactic space ($E \gapprox E_{knee})$. 

If we interpret a cosmic-ray energy spectrum as a size distribution
of particle energies, we can apply our universal SOC model. 
The driver of the SOC system is a generation mechanism of seed 
populations of charged particles, which are mostly bound to 
astrophysical objects in a collisional plasma. A critical threshold 
is given by transitions of the particles from collisional to 
collisionless plamsa (such as in the ``run-away regime''), 
where a particle can freely be accelerated, 
either by Fermi first-order or diffusive shock acceleration. The 
subsequent particle transport combined with numerous acceleration 
steps during every passage of suitable electric fields or shock 
fronts represents the build-up of an avalanche, until the particle 
hits the Earth's upper atmosphere where it is detected by a shower 
of secondary particles. If we could observe all end products of
an avalanche, we would expect an energy spectrum of $\alpha_E=1.5$.
In reality, the energy spectrum of cosmic rays is $\alpha_E \approx
3.0$, assuming that the detected energies are proportional to the
avalanche volume. How can we explain this discrepancy? A powerlaw
index of this value is expected for the size distribution of length
scales, $N(L) \propto L^{-3}$ (Eq.~1). Therefore, a similar energy spectrum
of $N(E) \propto E^{-3}$ can only be produced if the energy $E$ is
proportional to the length scale $L$, requiring that the fractal volume
$V_f(t) \propto r(t)^{D_3}$ has a fractal dimension of $D_3=1.0$. 
Such a scaling can be arranged if only a linear subvolume of an
entire 3D avalanche is observed, which is indeed the case for in-situ
detection at Earth, since the origin of the cosmic ray avalanche is
located far away. The situation is visualized in Fig.~8. 
In solar flares, on the other hand, almost all energetic particles 
accelerated during a flare lose their energy in the chromosphere,
and thus we can detect the entire energy content of an avalanche
event by remote-sensing. 
This is not possible in cosmic rays, because we cannot detect
in-situ all energy losses of cosmic ray particles that originated
isotropically from the same avalanche, in a remote place such as in
a supernova or black hole. Another aspect that our FD-SOC model 
predicts is the random walk diffusion during its propagation, which 
is consistent with the current thinking of cosmic ray particle transport.

Adopting the resulting scaling law between the energy $E$ of a
cosmic ray particle and the Euclidian length scale $L$ that a
cosmic ray particle has traveled at the time of detection,
$E \propto L$, which is expected for direct electric field
acceleration in a voltage drop, as well as for any other particle
acceleration mechanism with a fixed amount of energy extraction
per distance increase, $dE/dL \approx const$, we can even determine
the distance to the origination site of the cosmic ray particle.
The cosmic ray spectrum shown in Fig.~8 has a knee at 
$E_{free} \approx 0.5 \times 10^{16}$ eV, which marks the distance of the
Earth to the center of our galaxy $(L_{gal}\approx 50$ light years
or $L_{gal} \approx 5 \times 10^{22}$ cm). The Euclidean distance
where a cosmic ray particle with a maximum energy of $E_{max}
\approx 5 \times 10^{20}$ erg originated is then,
\begin{equation}
	L_{max} \approx L_{gal} \left( {E_{max} \over E_{knee}} \right) \ ,
\end{equation}
which yields $L_{max} \approx 5 \times 10^{27}$ cm, which corresponds
to about 10\% of the size of our universe ($R_{uni} \approx 4 
\times 10^{28}$ cm). Since the intergalactic and extragalactic
magnetic fields have different field strengths, the diffusion
coefficient of cosmic ray particles is also expected to be different
in these two regimes, which may explain the slightly different
powerlaw slopes below and above the galactic boundary $L_{gal}$
and the related energy $E_{knee}$. 

The lowest energies of the cosmic ray spectrum are at $E_{min} \approx
10^9$ eV. Using the same linear scaling of energy with length scale,
\begin{equation}
	L_{min} \approx L_{gal} \left( {E_{min} \over E_{knee}} \right) \ ,
\end{equation}
we estimate a distance of $L_{min} \approx 10^{16}$ cm or 200 AU, which
is located somewhat outside of the termination shock of our heliosphere.

\section{DISCUSSION}

In this paper we developed a macroscopic description of SOC systems
that is designed to reproduce the same statistical distributions of
observables as from SOC processes occurring on a microscopic level and
observed in nature. The microscopic processes cannot be treated
analytically due to the large number of degrees of freedom and the
nonlinear nature of the dynamic SOC systems. The complexity of
the microscopic fine structure during SOC avalanches is captured here 
in a approximative form by three simple parameters: the fractal 
dimension $D_d$, the diffusion coefficient $\kappa$,  and the 
diffusive spreading exponent $\beta$.  What is common to all
SOC processes is a system-wide critical threshold level that determines
whether ``avalanching'' occurs or not. For an overview we list the
physical mechanisms that operate in SOC systems in Table 2, containing 
a few classical SOC systems, as well as astrophysical applications 
that we described in this paper. We are aware that we use the term
``self-organized criticality'' in a more general sense than originally
envisioned, in the spirit of the definition given in the introduction: 
{\sl SOC is a critical state of a nonlinear energy dissipation system that 
is slowly and continuously driven towards a critical value of a system-wide
instability threshold, producing scale-free, fractal-diffusive, and 
intermittent avalanches with powerlaw-like size distributions.} 

Let us discuss the meaning of self-organizing criticality in astrophysical
applications in some more detail. Essentially we have three aspects of a
SOC system: (1) the energy input of the slow and steady driver,
(2) the self-organizing criticality condition or instability threshold,
and (3) the energy output in form of intermittent avalanches (Table 2).
The driver is necessary to keep a SOC process going, because the SOC
process would stop otherwise as soon as the system becomes subcritical.
The instability threshold $z_{crit}$ represents a bifurcation of two possible 
dynamic outcomes: either nothing happens when the state in every node of a
SOC system is below this critical threshold ($z < z_{crit}$), while an
avalanche or nonlinear energy dissipation event is triggered when a threshold 
exceeds at some location ($z \ge z_{crit}$). In classical SOC systems,
the self-organizing threshold can be a critical slope or angle of repose
(sandpile), a phase transition point (superconductor, Ising model, tea kettle), 
a fire ignition threshold (forest fires), or dynamic friction (earthquakes).
In astrophysical systems, the instability thresholds or critical values are
equally diverse, such as thresholds for magnetic instability with subsequent
magnetic reconnection (magnetospheric substorms, solar flares, stellar flares),
magnetic stressing (neutron star quakes, accretion disk flares),
particle acceleration thresholds, such as the ``run-away regime'' (solar
energetic particles, cosmic rays), vortex unpinning (neutron stars),
critical mass density for accretion (planetesimals, asteroids, accretion
disk flares, black-hole objects), gravitational disturbances and unstable
orbits that trigger collisions (Saturn ring particles, lunar craters), etc.
All these instability thresholds have system-wide critical values, so that
nothing happens below those values, while avalanching happens above their value. 
Since these instability thresholds or critical values occur system-wide,
set by physical conditions of the internal microscopic processes,
the system is self-organizing or self-tuning in the sense that it
maintains the same critical values throughout the system. This follows
the basic philosophy of Bak's sandpile, where a critical slope is
maintained system-wide, internally given by the critical value where
the gravitational and the dynamic friction forces are matching. 
For magnetic reconnection processes, for instance, critical values are
given by the kink-instability criterion or by the torus instability
criterion. For the formation of planetesimals (as well as for the
formation of planets and stars), a critical mass density is required 
where accretion by self-gravity overcomes diffusion. Thus, all these
astrophysical processes fulfill the basic requirement of our SOC 
definition, i.e., these nonlinear systems are slowly driven towards 
a critical value of a system-wide instability threshold. And all of these 
astrophysical processes exhibit scale-free, fractal-diffuse, and 
intermittent avalanches, with powerlaw-like size distributions.   

\section{CONCLUSIONS}

We can summarize the conclusions of this study as follows:

\begin{enumerate}
\item{We propose the following general definition of a SOC system:
{\sl SOC is a critical state of a nonlinear energy dissipation system that 
is slowly and continuously driven towards a critical value of a system-wide
instability threshold, producing scale-free, fractal-diffusive, and 
intermittent avalanches with powerlaw-like size distributions.} 
This generalized definition expands the original meaning of
self-tuning ``criticality'' to a wider class
of critical points and instability thresholds that have a similar 
(nonlinear) dynamical behavior and produce similar (powerlaw-like)
statistical size distributions.} 

\item{A macroscopic description of SOC systems has been derived from 
first principles that predicts powerlaw functions for the size
distributions of SOC parameters, as well as universal
values of the powerlaw slopes, for geometric and temporal parameters,
and some observables (flux and energy if they are proportional
to the emitting fractal volume). This macroscopic SOC model exhibits 
powerlaw scaling, universality, spatio-temporal correlations,
separation of time scales, fractality, and intermittency.
The predicted powerlaw slopes depend 
only on three parameters: on the Euclidean dimension $d$ of the 
system, the fractal dimension $D_d$, and the diffusive spreading 
exponent $\beta$. Note, that the spreading exponent is
an adjustable parameter in the FD-SOC model and can accomodate
classical diffusion, sub-diffusive, or hyper-diffusive transport,
and thus represents in some sense an ordering parameter, while
it cannot be adjusted in a branching process or in a BTW model
with a given re-distribution rule.}

\item{The FD-SOC model makes the following predictions: 
For the case of 3D space ($d=3$) and classical diffusion ($\beta=1$), 
the predicted values for the average fractal dimension is 
$D_d\approx (1+d)/3=2$, the powerlaw slopes are $\alpha_L=3$ 
for length scales, $\alpha_T=2$ for time scales, $\alpha_F=2$ 
for fluxes or energy dissipation rates, $\alpha_P=5/3$ for 
peak fluxes or peak energy dissipation rates, and $\alpha_E=3/2$ 
for fluences (i.e., time-integrated fluxes) or (total) avalanche energies,
assuming proportionality between the time-integrated fractal avalanche
volume and the observed fluence.}

\item{Among the astrophysical applications we find agreement 
between the predicted and observed size distributions for the
following phenomena: lunar craters, asteroid belts, Saturn ring
particles, auroral events during magnetospheric substorms, 
outer radiation belt electron bursts, solar flares,
soft gamma-ray repeaters, and blazars. This agreement between
theory and observations supports the universal applicability
of the fractal-diffusive SOC model.}

\item{Discrepancies between the predicted and observed size
distributions are found for stellar flares, pulsar glitches, 
black holes, and cosmic rays, but some can be reconciled with 
modified SOC models.  The disagreement for solar energetic particle 
(SEP) events is believed to be due to a
selection bias for large events. For stellar flares we conclude 
that the bolometric fluence is not proportional to the dissipated 
energy and volume. Pulsar glitches are subject to small-number statistics.
Black hole pulses have extremely steep size distributions that could
be explained by a suppression of large pulses for a certain period
after a large pulse. For cosmic rays, the energy size distribution
implies a fractal dimension of $D_3=1$ and a proportionality between
energy and length scales ($E \propto L$) according to our FD-SOC
model, which can be explained by the nature of in-situ detections
that capture only a small fraction of the avalanche volume.}
\end{enumerate}

Whatever the correct interpretations are for those phenomena 
with unexpected size distributions, the application of our
standard FD-SOC model can reveal alternative scaling laws
that can be tested in future measurements. A major achievement
of our standard FD-SOC model is the fact that it can predict 
and explain, in a universal way, the powerlaw indices of different 
SOC parameters (lengths, durations, fluxes, energies, waiting times)
in most of the considered astrophysical applications,
which do not depend on the details of the
underlying physical mechanisms. We have also to appreciate that
the macroscopic approach of SOC statistics does not depend on the
microscopic fine structure of each SOC process, unlike the 
mathematical/numerical SOC models, which produce different powerlaw
slopes depending on the assumed re-distribution rule, and partially
do not fulfill universality. Our macroscopic fractal-diffusive SOC
model may also be suitable to correctly describe the statistics of
other, SOC-related, nonlinear processes, such as percolation or 
turbulence, an aspect that needs to be investigated in future.

\acknowledgements
The author acknowledges the insightful and helpful comments by the
referee, and thanks Yuta Notsu and Takuya Shibayama for providing
Kepler data. This work has benefitted from fruitful discussions with 
Karel Schrijver, Henrik Jensen, Nicholas Watkins, J\"urgen Kurths,
Vadim Uritsky,
and by the {\sl International Space Science Institute (ISSI)}
at Bern Switzerland, which hosted and supported two workshops on
{\sl Self-Organized Criticality and Turbulence} during October
15-19, 2012 and September 16-20, 2013.  This work was partially 
supported by NASA contract NNX11A099G ``Self-organized criticality 
in solar physics'' and NASA contract NNG04EA00C of the SDO/AIA 
instrument to LMSAL.

%%%%%%%%%%%%%%%%%%%%%%%%%%%%%%%%%%%%%%%%%%%%%%%%%%%%%%%%%%%%%%%%%%%%%%%

\begin{table}[t]
\begin{center}
\normalsize
\caption{Summary of theoretically predicted and observed powerlaw
indices of size distributions in astrophysical systems.}
\medskip
\begin{tabular}{|l|l|l|l|l|l|l|}
\hline
                                 & Length     & Area        & Duration    & Peak flux   & Energy     & Waiting \\
                                 & $\alpha_L$ & $\alpha_A$,
                                                $\alpha_{th,A}$ & $\alpha_T$  & $\alpha_P$  &$\alpha_E$ &time $\alpha_{\Delta t}$\\
\hline
FD-SOC Theory                    & {\bf 3.0}  & {\bf 2.0}   & {\bf 2.0}   & {\bf 1.67}  & {\bf 1.5} & {\bf 2.0} \\
\hline
\underbar{Lunar craters:}        &            &             &             &             &            & \\
Mare Tranquillitatis $^1)$       & 3.0        &             &             &             &            & \\
Meteorites and debris $^2)$      & 2.75       &             &             &             &            & \\
\hline
\underbar{Asteroid belt:}        &            &             &             &             &            & \\
{\sl Spacewatch Surveys}$^3)$    & 2.8        &             &             &             &            & \\
{\sl Sloan Survey}$^4)$          & 2.3-4.0    &             &             &             &            & \\
{\sl Subaru Survey}$^5)$         & 2.3        &             &             &             &            & \\
\hline
\underbar{Saturn ring:}          &            &             &             &             &            & \\
Voyager 1$^6)$                   & 2.74-3.11  &             &             &             &            & \\
\hline
\underbar{Magnetosphere:}        &            &             &             &             &            & \\
EUV auroral events$^7$           &            &1.73$-$1.92  &2.08$-$2.39  &1.66$-$1.82  &1.39$-$1.61 & \\
Optical auroral events$^8$       &            &1.85$-$1.98  &2.25$-$2.53  &1.71$-$2.02  &1.50$-$1.74 & \\
Outer radiation belt$^{9})$      &            &             &             &1.5$-$2.1    &            & \\
\hline
\underbar{Solar Flares:}         &            &             &             &             &            & \\
HXR, ISEE-3$^{10}$               &            &             & 1.88-2.73   & 1.75-1.86   & 1.51-1.62  & \\
HXR, HXRBS/SMM$^{11}$            &            &             &$2.17\pm0.05$&$1.73\pm0.01$&$1.53\pm0.02$&2.0$^a$ \\
HXR, BATSE/CGRO$^{12}$           &            &             & 2.20-2.42   & 1.67-1.69   & 1.56-1.58  & 2.14$\pm$0.01$^b$ \\
HXR, RHESSI$^{13}$               &            &             & 1.8-2.2     & 1.58-1.77   & 1.65-1.77  & 2.0$^a$ \\
SXR, Yohkoh$^{14}$               &1.96$-$2.41 & 1.77-1.94   &             & 1.64-1.89   & 1.4-1.6    & \\
SXR, GOES$^{15}$                 &            &             & 2.0-5.0     & 1.86-1.98   & 1.88       & 1.8$-$2.4$^c$ \\
EUV, SOHO/EIT$^{16}$             &            & 2.3-2.6     & 1.4-2.0     &             &            & \\
EUV, TRACE$^{17}$                &2.50$-$2.75 & 2.4-2.6     &             & 1.52-2.35   & 1.41-2.06  & \\
EUV, AIA/SDO$^{18}$              &3.2$\pm$0.7 &$2.1\pm0.3$  &$2.10\pm0.18$&$2.0\pm0.1$  &$1.6\pm0.2$ & \\
EUV, EIT/SOHO$^{19}$             &3.15$\pm$0.18&2.52$\pm$0.05&1.79$\pm$0.03&            &1.48$\pm$0.03 & \\
Magnetic events, MDI/SOHO$^{19}$ &2.57$\pm$0.13&1.93$\pm$0.06&2.02$\pm$0.07&            &1.47$\pm$0.03 & \\
Radio microwave bursts$^{20}$    &            &             &             & 1.2-2.5     &            & \\
Radio type III bursts$^{21}$     &            &             &             & 1.26-1.91   &            & \\
Solar energetic particles$^{22}$ &            &             &             & 1.10-2.42   & 1.27-1.32  & \\
\hline
\underbar{Stellar Flares:}       &            &             &             &             &            & \\
EUVE flare stars$^{23}$          &            &             &             &             &$2.17\pm0.25$& \\
KEPLER flare stars$^{24}$        &            &             &             &$1.88\pm0.09$&$2.04\pm0.13$& \\
\hline
\underbar{Astrophysical Objects:}&            &             &             &             &            & \\
Crab pulsar$^{25}$               &            &             &             & 3.06-3.50   &            & \\
PSR B1937+21$^{26}$              &            &             &             & $2.8\pm0.1$ &            & \\
Soft Gamma-Ray repeaters$^{27}$  &            &             &             &             &$1.43-1.76$ & \\
Cygnus X-1 black hole$^{28}$     &            &             &             & 7.1         &            & \\
Blazar GC 0109+224$^{29}$        &            &             &             & 1.55        &            & \\
Cosmic rays$^{30}$               &            &             &             &             &$2.7-3.3$   & \\
\hline
\end{tabular}
\end{center}
\end{table}

\clearpage
{ %\footnotesize 
\underbar{References to Table 13.2:}
$^1)$ Cross (1966);
$^2)$ Sornette (2004);
$^3)$ Jedicke and Metcalfe (1998);
$^4)$ Ivezic et al.~(2001);
$^5)$ Yoshida et al.~(2003), Yoshida and Nakamura (2007);
$^6)$ Zebker et al.~(1985), French and Nicholson (2000);
$^7)$ Uritsky et al.~(2002);
$^8)$ Kozelov et al.~(2004);
$^{9})$ Crosby et al.~(2005)
$^{10})$ Lu et al.~(1993), Lee et al.~(1993);
$^{11})$ Crosby et al.~(1993);
$^{12})$ Aschwanden ~(2012a,2011b);
$^{13})$ Christe et al.~(2008), Lin et al.~(2001), Aschwanden ~(2011a,c);
$^{14})$ Shimizu (1995), Aschwanden and Parnell~(2002);
$^{15})$ Lee et al.~(1995), Feldman et al.~(1997), Veronig et al.~(2002a,b),
         Aschwanden and Freeland (2012);
$^{16})$ Krucker and Benz (1998), McIntosh and Gurman (2005);
$^{17})$ Parnell and Jupp (2000), Aschwanden et al. 2000, Benz and Krucker (2002),
         Aschwanden and Parnell (2002), Georgoulis et al.~(2002);
$^{18})$ Aschwanden and Shimizu (2013), Aschwanden, Zhang, and Liu (2013); 
$^{19}$ Uritsky et al.~(2002);
$^{20})$ Akabane (1956), Kundu (1965), Kakinuma et al.~(1969), 
	 Das et al.~(1997), Nita et al.~(2002);
$^{21})$ Fitzenreiter et al.~(1976), Aschwanden et al.~(1995), Das et al.~(1997), Nita et al.~(2002);
$^{22})$ Van Hollebeke et al.~(1975), Belovsky and Ochelkov (1979), Cliver et al. (1991),
         Gabriel and Feynman (1996), Smart and Shea (1997), Mendoza et al.~(1997),
         Miroshnichenko et al.~(2001), Gerontidou et al.~(2002);
         Gabriel and Feynman (1996);
$^{23})$ Robinson et al.~(1999). Audard et al.~(2000), Kashyap et al.~(2002),
         G\"udel et al.~(2003), Arzner and G\"udel (2004), Arzner et al.~(2007), 
	 Stelzer et al.~(2007);
$^{24})$  Maehara et al. 2012; Shibayama et al.~2013;
$^{25})$ Argyle and Gower (1972), Lundgren et al.~(1995);
$^{26})$ Cognard et al.~(1996);
$^{27})$ Gogus et al.~(1999, 2000);
$^{28})$ Negoro et al.~(1995), Mineshige and Negoro (1999);
$^{29})$ Ciprini et al.~(2003);
$^{30})$ e.g., Fig.~13.18 (courtesy of Simon Swordy, Univ.Chicago);
$^{a}) $ Aschwanden and McTiernan (2010);
$^{b}) $ Grigolini et al.~(2002);
$^{c}) $ Wheatland (2001, 2003), Boffetta et al.~(1999), Lepreti et al.~(2001).}

\clearpage

%%%%%%%%%%%%%%%%%%%%%%%%%%%% TABLE 2 %%%%%%%%%%%%%%%%%%%%%%%%%%%%%%%%%%%%
\begin{table}[t]
\begin{center}
\small
\caption{Physical mechanisms operating in self-organized criticality systems.}
\medskip
\begin{tabular}{|l|l|l|l|}
\hline
Phenomenon	& Energy Input 		& Instability threshold & Energy output   \\
		& (steady driver)	& (criticality)	   & (intermittent avalanches) \\	
\hline
CLASSICAL SOC: 	&			&		   &		     \\	
Sandpile 	& gravity (dripping sand)& angle of repose & sand avalanches \\
Superconductor	& magnetic field change & phase transition & vortex avalanches\\
Ising model	& temperature increase  & phase transition & atomic spin-flip\\ 
Tea kettle	& temperature increase  & boiling point	   & vapour bubbles  \\	
Earthquakes	& tectonic stressing	& dynamical friction & rupture area    \\
Forest fire	& tree growth           & fire ignition point & burned area     \\
BTW cellular automaton & input at random nodes& critical threshold &next-neighbor redistribution \\ 
\hline
ASTROPHYSICS:	&			&		   &		     \\	
Lunar craters	& meteorite production	& lunar collision  & lunar impact craters\\
Asteroid belt	& planetesimals		& critical mass density & asteroids       \\
Saturn ring	& gravitational disturbances & collision rate  & Saturn ring particles\\
Magnetospheric substorm& solar wind	& magnetic reconnection & auroral bursts \\
Radiation belt  & solar wind            & magnetic trapping/untrapping & electron bursts	\\
Solar flares	& magnetic stressing	& magnetic reconnection & nonthermal particles\\
Stellar flares	& magnetic stressing	& magnetic reconnection & nonthermal particles\\
Pulsar glitches & neutron star spin-up  & vortex unpinning  &neutron starquakes\\
Soft gamma-ray repeaters& magnetic stressing &star crust fracture &neutron starquakes\\
Black-hole objects & gravity            & accretion and inflow   &X-ray bremsstrahlung pulses\\ 
Blazars		& quasar jets		& jet direction jitter & optical radiation pulses\\
Cosmic rays	& galactic magnetic fields& (run-away) acceleration threshold & high-energy particles\\
\hline
\end{tabular}
\end{center}
\end{table}

%%%%%%%%%%%%%%%%%%%%%%%%%%% FIGURE %%%%%%%%%%%%%%%%%%%%%%%%%%%%%%%%% 

\begin{figure}
\plotone{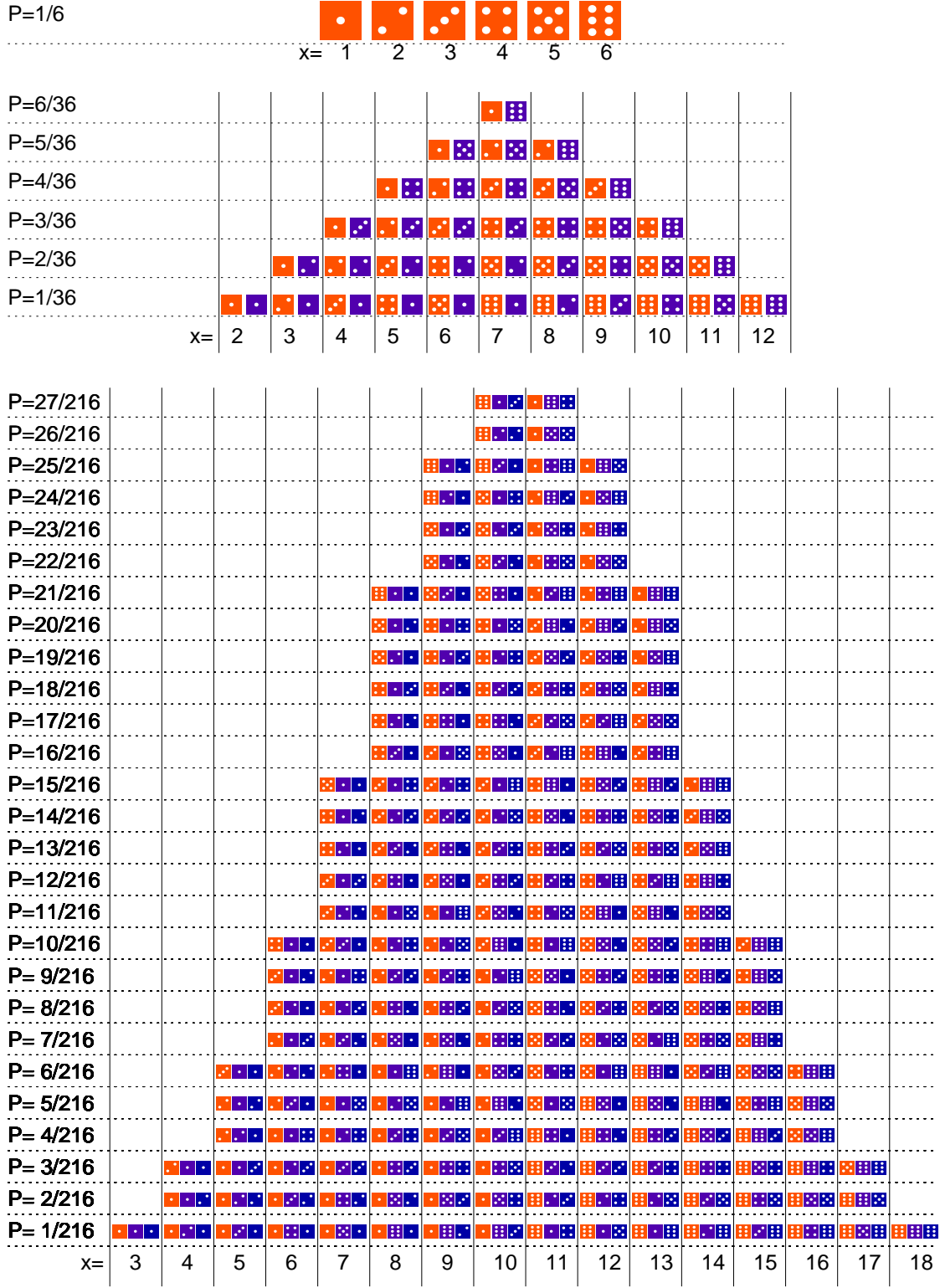}
\caption{The probability distribution $P(x)$ of statistical outcomes $x$
are shown for tossing one (top panel), two (second panel), and three
dice (bottom panel). The possible outcomes cover the ranges of
$x=x_1=1,...,6$ for one dice, $x=(x_1+x2)=2,...,12$ for two dice, and
$x=(x_1+x_2+x_3)=3,...,18$ for three dice. The probability distributions
are also known as binomial distributions and converge to a Gaussian
distribution for an infinite number of dice.}
\end{figure}

\begin{figure}
\plotone{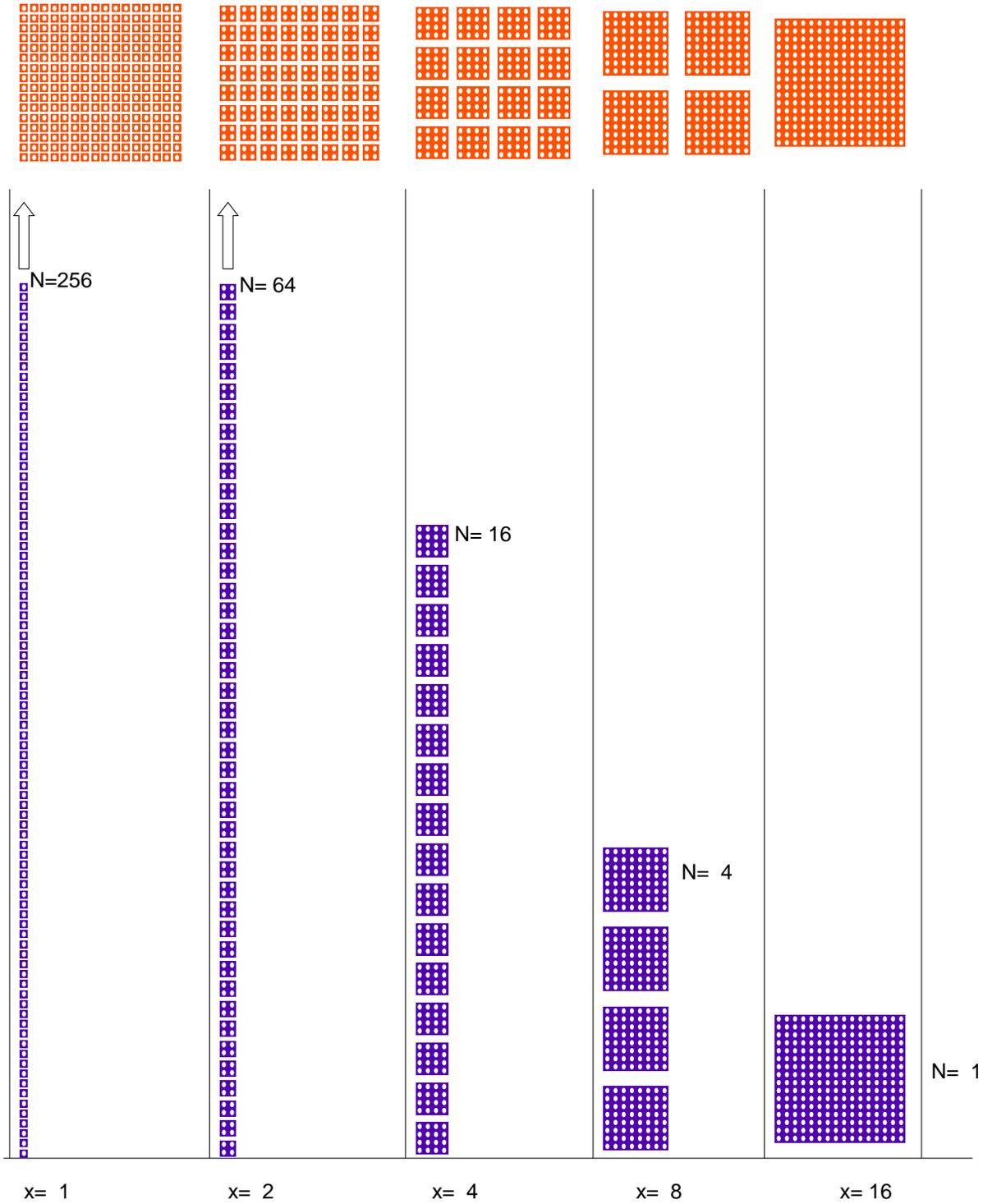}
\caption{The probability distributions $N(x)$ of statistical outcomes
$x$ are shown for braking domino pieces with square-like shape into
smaller squares with side lengths that correspond to powers of two
(i.e., x=1,2,4,8,16). A histogram with such logarithmic bins shows
the number of outcomes, $N(x=1)=256=2^5$,
$N(x=2)=64=2^4$, ...., $N(x=16)=1$, which form a powerlaw distribution
$N(x) \propto x^{-2}$ with a slope of $\alpha=\log{(N)}/\log{(x)}=-2$.}
\end{figure}

\begin{figure}
\plotone{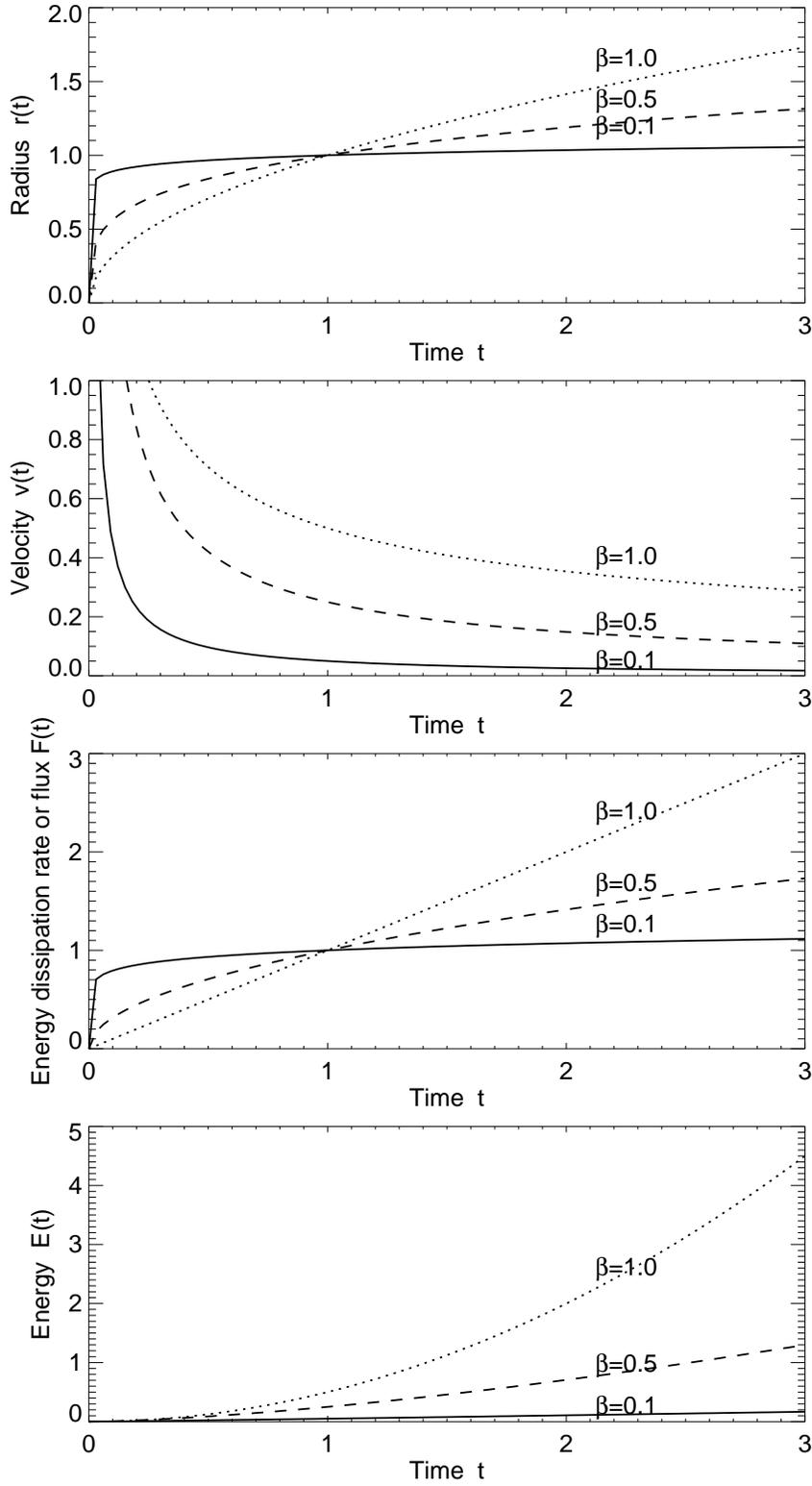}
\caption{The spatio-temporal evolution of the avalanche radius $r(t)$
is shown (top panel), the expansion velocity $v(t)$ (second panel),
the energy dissipation rate or flux $F(t)$ (third panel),
and the dissipated energy $E(t)$ (bottom panel), 
are shown for our macroscopic FD-SOC model for
diffusive spreading exponents of $\beta=0.1$ (quasi-logistic; solid
line style), $\beta=0.5$ (sub-diffusive; dashed line style), and 
$\beta=1$ (classical diffusion, solid line style).}
\end{figure}

\begin{figure}
\plotone{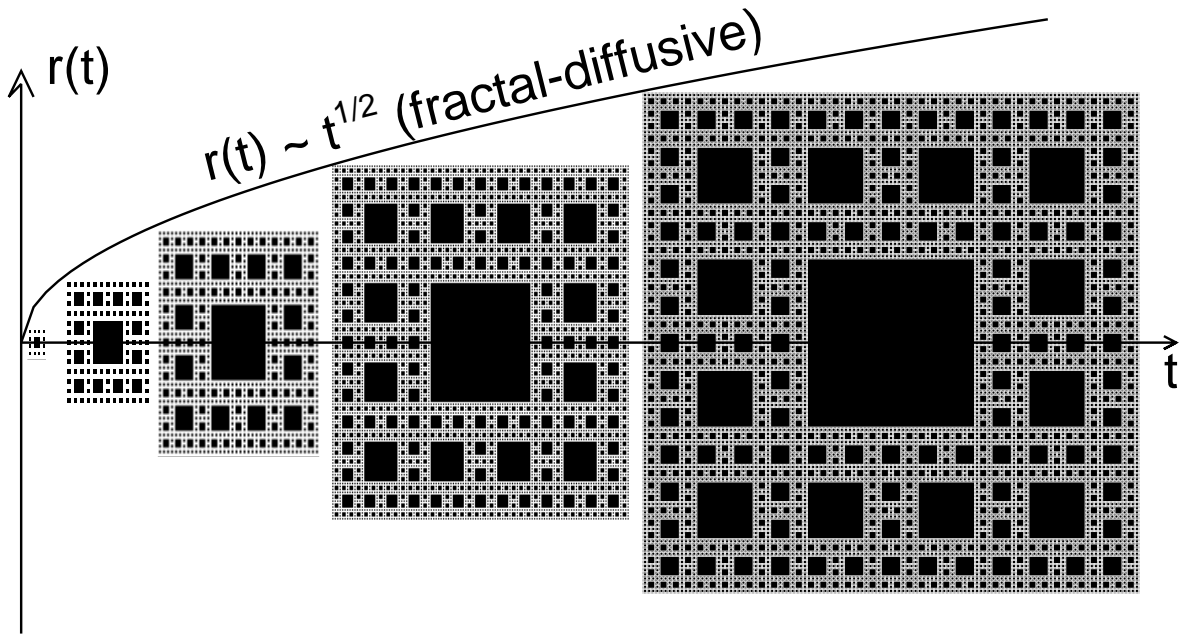}
\caption{A cartoon that illustrates the concept of fractal-diffusive
avalanche evolution. The Euclidean radius $r(t)$ evolves like a 
diffusive random walk, such as $r(t) \propto t^{1/2}$ for classical 
diffusion, while the avalanche area is fractal (black substructures). 
The instantaneous fractal area $A_f(t) \propto r(t)^{D_t}$ consists 
of the number active nodes and is proportional to the energy dissipation 
rate $dE(t)/dt$ or flux $F(t)$ at a given time $t$.}
\end{figure}

\begin{figure}
\plotone{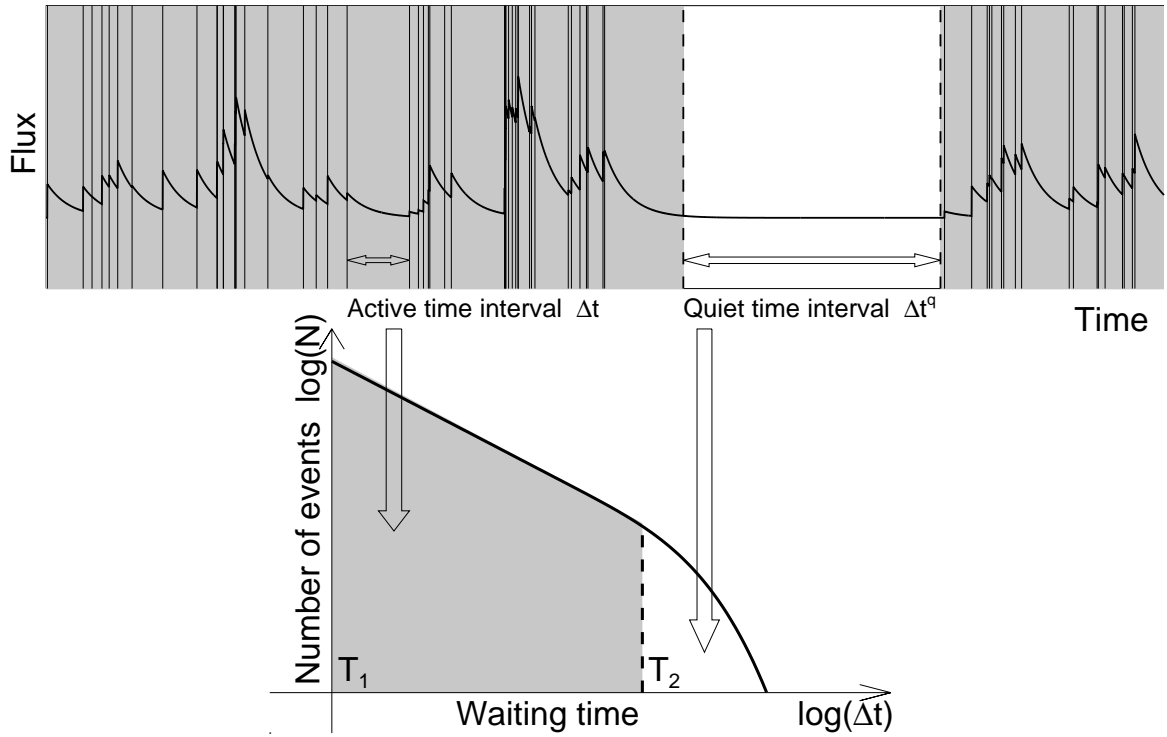}
\caption{The concept of a dual waiting time distribution is illustrated,
consisting of active time intervals $\Delta t \lapprox T_2$) that contribute
to a powerlaw distribution (which is equal to that of time durations, $N(T)$),
and random-like quiet time intervals ($\Delta t^q$) that contribute to an 
exponential cutoff.}
\end{figure}

\begin{figure}
\plotone{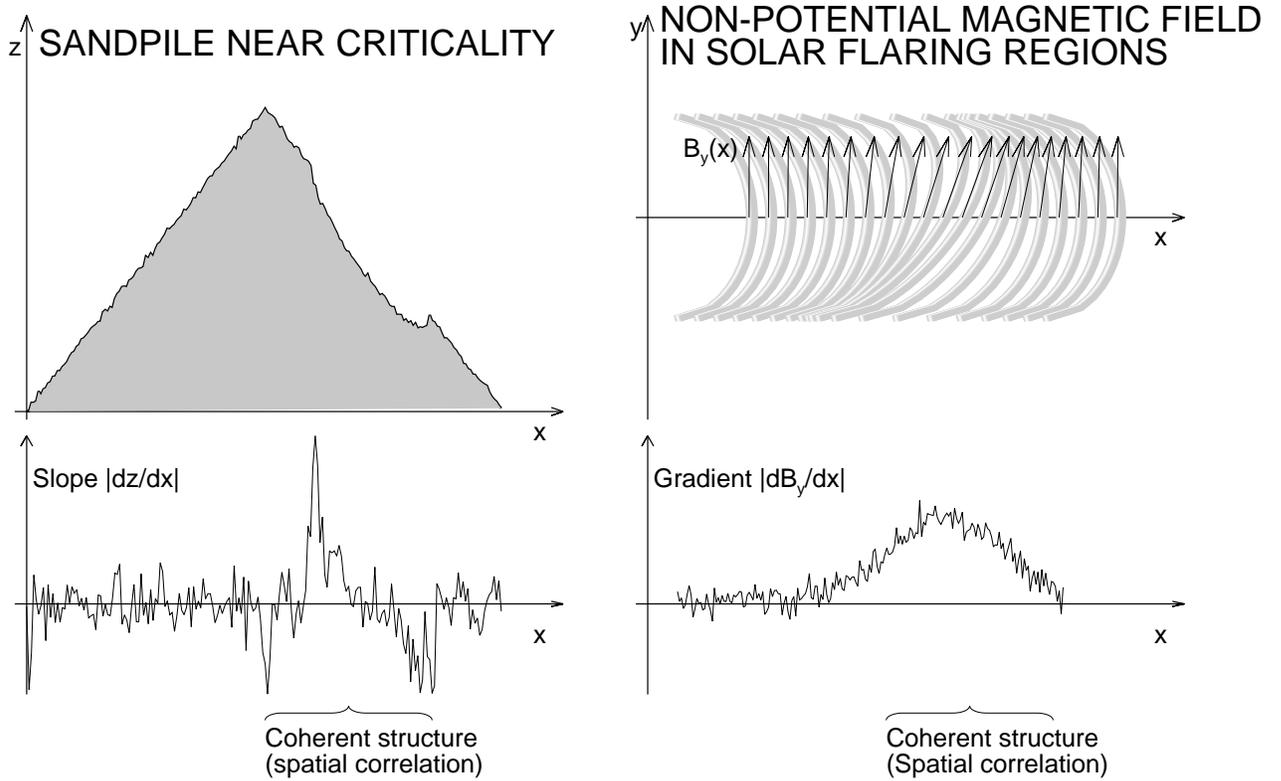}
\caption{{\sl Left:} A sandpile in a state in the vicinity of criticality
is shown with a vertical cross-section $z(x)$, with the gradient of the slope
(or repose angle) $|dz/dx|$ (bottom), exhibiting short-range fluctuations
due to noise and long-range correlation lengths due to locally extended
deviations
from the mean critical slope. {\sl Right:} The solar analogy of a flaring
region is visualized in terms of a loop arcade over a neutral line
in $x$-direction, consisting of loops with various shear angles that
are proportional to the gradient of the field direction $B_x/B_y$,
showing also some locally extended (non-potential) deviations from
the potential field (bottom).}
\end{figure}

\begin{figure}
\plotone{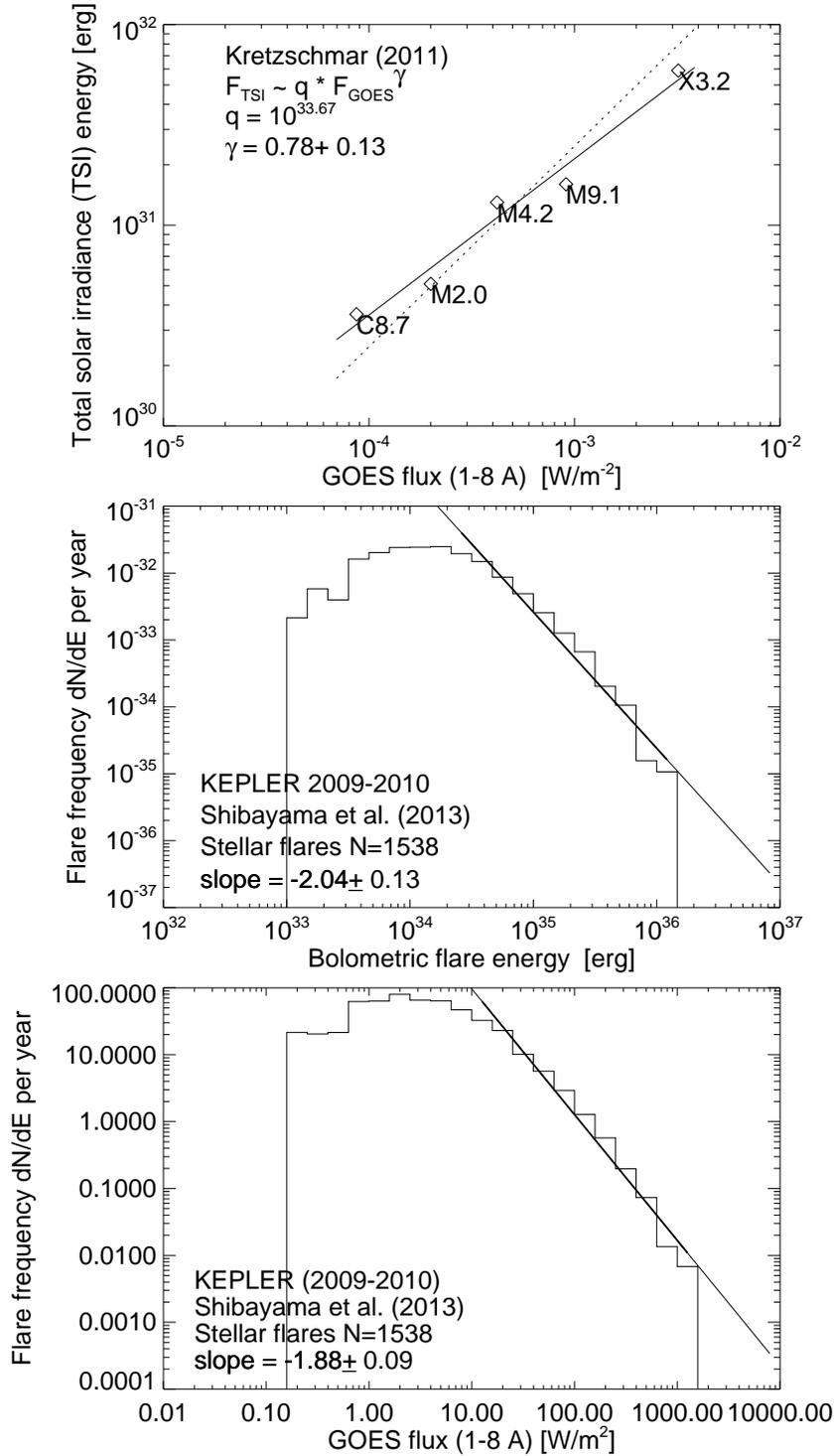}
\caption{The scaling law of the total solar irradiance (TSI) and
the GOES 1-8 \ang\ peak flux based on a linear regression fit (solid line)
to data from Kretzschmar (2011) is shown, i.e., $E_b \propto 
P_x^{0.78\pm0.13}$ (top panel). A linear relationship is indicated
with a dotted line. The bolometric flare energy of 1538 stellar flares observed
with {\sl Kepler} is histogrammed, yielding a size distribution with a
powerlaw slope of $\alpha_E = 2.04\pm0.13$ (middle panel), and the
inferred size distribution of GOES fluxes using the scaling law of
Kretzschmar (2011), yielding a powerlaw slope of $\alpha_P=1.88\pm0.09$
(bottom panel).}
\end{figure}

\begin{figure}
\plotone{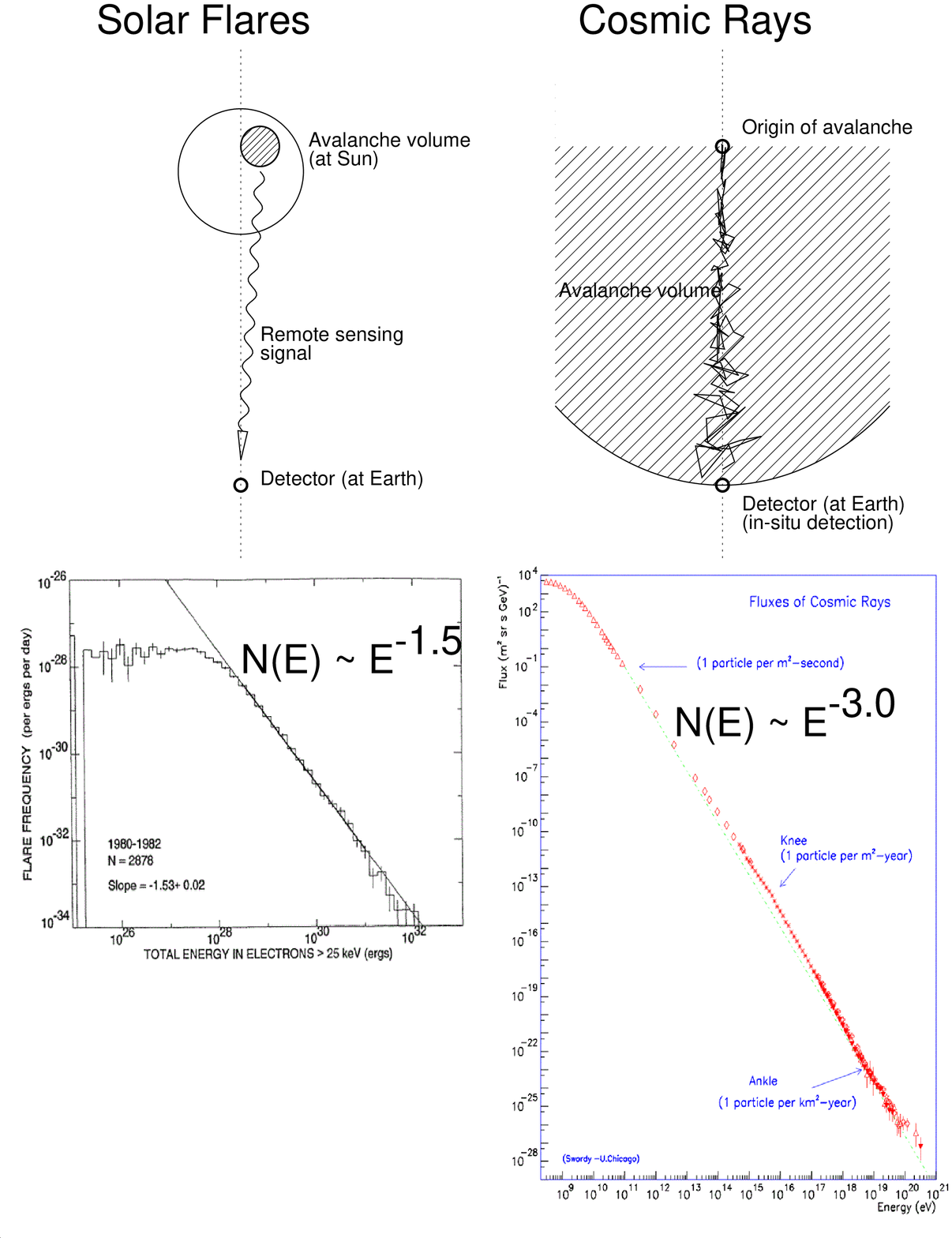}
\caption{The energy of solar flares is proportional to the avalanche
volume, which is probed in its entire 3D volume by remote sensing
via soft or hard X-rays (top left), while the energy of cosmic
rays detected in-situ is inferred from a linear sub-volume of the entire
3D avalanche only (top right). The predicted energy size distributions
are therefore different, with $N(E) \propto E^{-1.5}$ based on the 
3D Euclidean volume for solar flares, (bottom left; Crosby et al.~1993), 
and $N(E) \propto E^{-3.0}$ based on a 1D sub-volume for cosmic rays
(bottom right; Credit: Simon Swordy, University of Chicago).}
\end{figure}

\end{document}